\documentclass[aps,prd,noeprint,showpacs,twocolumn,amsmath,amssymb,superscriptaddress,nofootinbib]{revtex4-2}

\usepackage[normalem]{ulem}
\usepackage{float}
\usepackage[varg]{txfonts}
\usepackage[]{graphicx}
\usepackage{enumerate}
\usepackage{subeqnarray}
\usepackage{cases}
\usepackage{mathrsfs}
\usepackage[usenames]{color}
\usepackage{bibentry}
\usepackage{hyperref}
\hypersetup{colorlinks=true, citecolor=blue, urlcolor=blue, linkcolor=blue}

\providecommand{\aap}{Astron. Astrophys.}
\providecommand{\mnras}{Mon. Not. R. Astron Soc.}
\providecommand{\apjl}{Astrophys. J. Lett}
\providecommand{\ssr}{Space Sci. Rev.}

\providecommand{\apjs}{Astrophys. J. Suppl. Ser.}
\providecommand{\araa}{Annu. Rev. Astron. Astrophys.}
\providecommand{\physrep}{Phys. Rep.}
\providecommand{\aapr}{Astron. Astrophys. Rev.}
\providecommand{\csr}{Chem. Soc. Rev.}

\begin{document} 
\author{D.~O.~Chernyshov}
\email{chernyshov@lpi.ru}
\affiliation{I.~E.~Tamm Theoretical Physics Division of P.~N.~Lebedev Institute of Physics, 119991 Moscow, Russia}
\author{A.~V.~Ivlev}
\email{ivlev@mpe.mpg.de}
\affiliation{Max-Planck-Institut f\"ur extraterrestrische Physik, 85748 Garching, Germany}
\author{V.~A.~Dogiel}
\affiliation{I.~E.~Tamm Theoretical Physics Division of P.~N.~Lebedev Institute of Physics, 119991 Moscow, Russia}

\title{Self-consistent model of cosmic ray penetration into molecular clouds:\\ Effect of energy losses}


\date{\today}
\setcitestyle{authoryear,round}
\begin{abstract}
The theory of cosmic-ray (CR) penetration into dense molecular clouds developed recently for relativistic particles by \citet{Chernyshov2024} is extended to non-relativistic CRs. Interstellar CRs streaming into the clouds are able to resonantly excite MHD waves in diffuse cloud envelopes. This leads to the self-modulation, such that streaming particles are scattered at the self-generated waves. In contrast to relativistic CRs, transport of lower-energy particles in the envelopes is generally heavily affected by ionization losses; furthermore, both CR protons and electrons contribute to wave excitation. We show that these effects have profound impact on the self-modulation, and can dramatically reduce CR spectra even for clouds with moderate column densities of a few times $10^{21}$ cm$^{-2}$.
\end{abstract}
\setcitestyle{numbers,square}


   
   \maketitle

\section{Introduction}
\label{intro}

Interstellar (ISM) cosmic rays (CRs) are able to resonantly excite MHD waves in diffuse gas surrounding dense molecular clouds while streaming into the clouds \citep{skill76, ivlev18, dogiel2018}. This phenomenon
can lead to {\it self-modulation} of penetrating CRs: streaming particles are efficiently scattered at the self-generated waves and, as a result, the CR spectrum in the cloud is reduced compared to the ISM spectrum. 

\begin{figure}
    \centering
    \includegraphics[width=\linewidth]{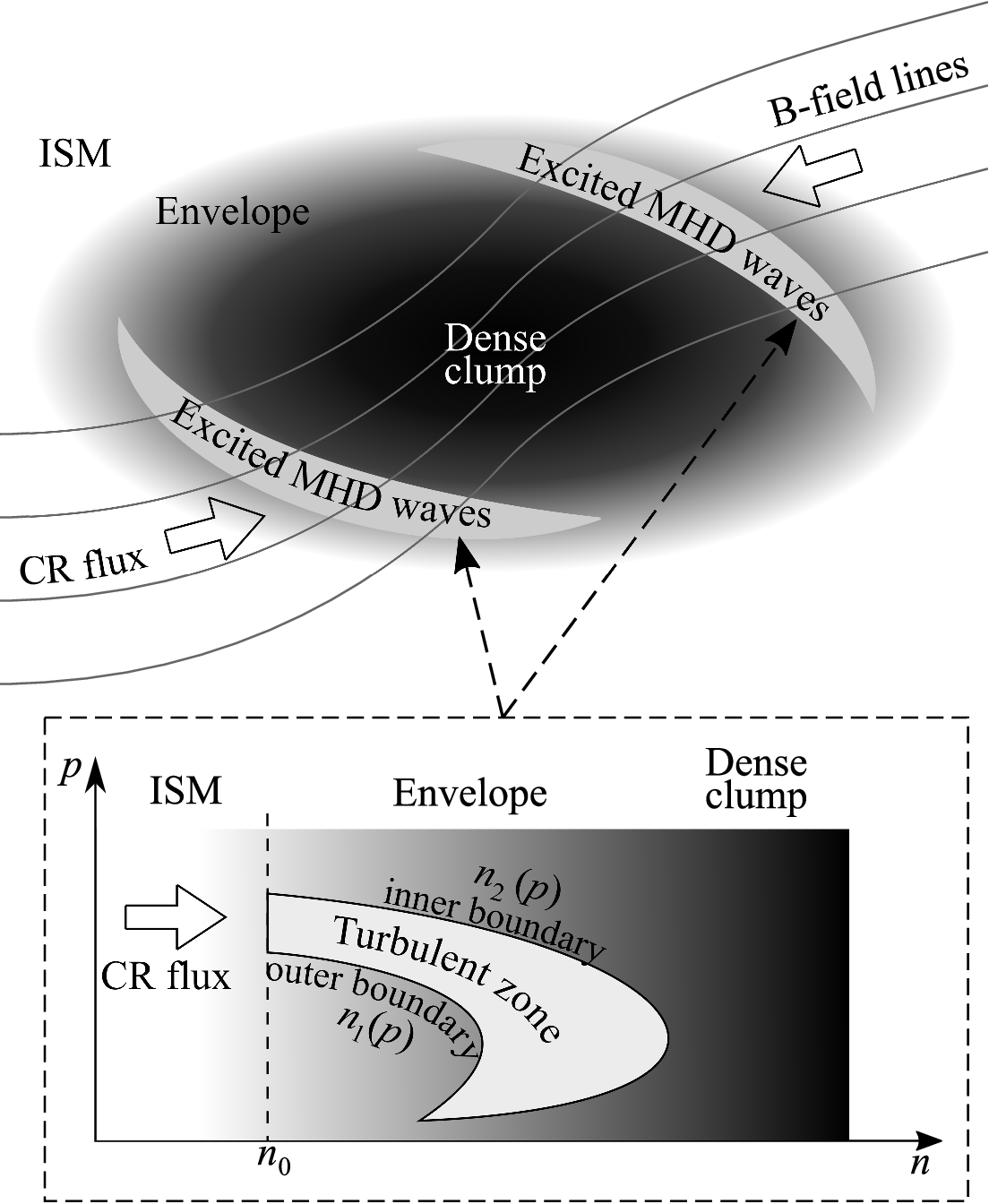}
    \caption{A sketch illustrating self-modulation of CRs penetrating a dense molecular cloud. Interstellar CRs, being isotropic in the absence of the cloud, stream along the local magnetic field lines in both directions. The net flux of CRs into the cloud is formed due to their attenuation in the central dense clump. This triggers the resonant streaming instability in the diffuse envelope surrounding the clump, generating the turbulent zones in outer envelope regions. The gas density $n$ in the envelope increases monotonically toward the dense clump; the inset highlights the fact that the outer and inner boundaries of the turbulent zone, $n_1(p)$ and $n_2(p)$, are functions of the particle momentum $p$.  
    }
    \label{fig:sketch}
\end{figure}

An overall concept of CR self-modulation, discussed in detail in Refs.~\citep{Chernyshov2024} and \citep{ivlev18}, is sketched in Fig.~\ref{fig:sketch}. Consider a molecular cloud with a dense sub-parsec clump in the center (completely dominating the total column density of the cloud) surrounded by diffuse parsec-scale envelope. Interstellar CRs entering on either side of the cloud along the local magnetic field lines lose a certain fraction of their energy while crossing the clump, which leads to a net flux of CRs into the cloud. The flux velocity, an increasing function of the cloud column density \citep{dogiel2018}, may exceed the local velocity of Alfven waves -- which is the necessary condition for the resonant streaming instability \citep{kuls69}. If the excitation rate due to free-streaming CRs exceeds the wave damping rate, the locally generated turbulence starts scattering penetrating CRs, such that some particles are reflected back to the ISM. The resulting CR flux is then self-regulated by the condition that the wave excitation is exactly balanced by damping. As the damping rate is proportional to the local gas density, the wave excitation occurs in outer regions of diffuse envelope, near the transition to the ISM.

In our recent paper by \citet{Chernyshov2024} we studied self-modulation of relativistic CRs. We generalized our earlier model by \citet{ivlev18} and \citet{dogiel2018} (in which the value for the envelope's gas density where CRs are able to excite waves was treated as a free parameter) and obtained a universal analytical solution, applicable for arbitrary density distribution monotonically increasing in the envelope toward the dense clump. In Ref.~\citep{Chernyshov2024}, we
self-consistently derived the CR diffusion coefficient as well as the outer and inner boundaries of the {\it turbulent zone} (or {\it diffusion zone}), the region where the wave excitation takes place. Location of these boundaries depends on the particle momentum (energy), as depicted in the inset of Fig.~\ref{fig:sketch}. Waves can only be excited within the envelope, since the damping rate is too high in the warm neutral ISM \citep{Chernyshov2024}.


The model by \citet{Chernyshov2024} was tested by computing the impact of CR self-modulation on the gamma-ray emission. We obtained excellent quantitative agreement with the recent Fermi LAT observations of nearby giant molecular clouds \citep{yang2023}, showing characteristic deficits in the emission at energies below a few GeV, as predicted by the theory.


The aim of the present paper is to extend the results of \citet{Chernyshov2024} to non-relativistic CRs. This work is motivated by the fact that the low-energy part of the interstellar CR spectrum determines ionization and thus plays a crucial role in the evolution of molecular clouds, controlling multiple physical and chemical processes that accompany practically all stages of star formation (see, e.g., reviews by \citet{Padovani2020, Gabici2022}, and references therein). Our research will help in providing more insights into the potential sources and transport models of low-energy CRs, since such particles also determine the 6.4~keV Fe~K$\alpha$ line emission as well as the nuclear de-excitation line emission from molecular gas \cite{Ramaty79, Tatischeff03, Benhabiles2013, Phan2020, Fujita2021, Liu2021, shi2024}. 

We show that the major difference with respect to relativistic particles, whose propagation in diffuse envelopes is loss free \citep{Chernyshov2024}, is that (i) ionization losses now become generally important, and (ii) both CR protons and electrons contribute to wave excitation. Our analysis suggests that these effects can dramatically reduce spectra of penetrating CRs even for clouds with moderate column densities.

\section{Governing equations and the solution}
\label{gov_eqs}

According to \citet{Chernyshov2024} and \citet{ivlev18}, the CR streaming instability produces a turbulent zone, the region where penetrating particles are efficiently scattered at the self-generated waves. The turbulent zone has a crescent shape in the plane spanned by the CR energy/momentum and the gas density (see the inset in Fig.~\ref{fig:sketch}), with the upper tip set by the {\it excitation threshold} -- the maximum energy at which the CR flux is able to trigger waves. Depending on the interstellar spectrum and the cloud column density, the CR spectrum in the cloud interior may be significantly modulated at energies below the excitation threshold \citep{Chernyshov2024}.

Outer regions of diffuse envelopes have negligible contribution to the total column density of molecular clouds. This implies that the net flux velocity of penetrating CRs is completely determined by their attenuation in the dense clump \citep{Chernyshov2024}. At the same time, generation of waves is only possible in diffuse envelopes, as the waves are efficiently damped at higher densities. Thus, the solution for the turbulent zone is reduced to analysis of the processes occurring in the envelope, while the dense clump only sets the value of the flux velocity (see Sec.~\ref{BCs}).

Propagation of CRs within the turbulent zone is generally described by the advection-diffusion transport equation,
\begin{equation}
\frac{\partial}{\partial z}\left( v_{\rm A}f - D\frac{\partial f}{\partial z}\right) - \frac{\partial}{\partial p}\left(\dot{p}f\right) = 0\,,
\label{eq:main_propagation23}
\end{equation}
where $f(p,z)$ is the CR distribution function (spectrum) in the momentum space, normalized such that the local number density of CRs is $\int f(p,z) dp$, and $z$ is the coordinate along the magnetic field line. The flux, $v_{\rm A}f - D\partial f/\partial z\equiv S(p,z)$, is a sum of the advection and diffusion components: the latter is proportional to the diffusion coefficient $D(p)$, the former is set by the velocity of self-excited waves, equal to the Alfven velocity $v_{\rm A}(z) = B/\sqrt{4\pi m_i \xi_i n}\,$, which is determined by the longitudinal magnetic field $B(z)$, the gas density $n(z)$, and the mass $m_i$ of the dominant ions with the abundance $\xi_i$. 

Attenuation of CRs, controlled by continuous losses due to interaction with gas, is described by the momentum loss function $\dot{p}>0$. The dominant loss mechanism for non-relativistic particles is ionization. At higher energies, losses become catastrophic: they are determined by the pion production (fragmentation) for protons (nuclei), and by bremsstrahlung for electrons. As shown in Ref.~\citep{Padovani2018} (see their Fig.~1 and Appendix~A.2), synchrotron losses are typically negligible for electron energies well below $\sim 1$~TeV. 
Finally, contribution of adiabatic losses, representing conversion of the CR energy into the energy of MHD waves \citep{ivlev18, BerezinskiiBook1990}, can also be neglected in our problem (see Sec.~\ref{loss_effect}).

The rate of resonant wave excitation by streaming CRs is proportional to the diffusion component of the flux \citep{Skilling1975}. Waves are damped due to ion-neutral collisions, and therefore the diffusion term in Eq.~(\ref{eq:main_propagation23}) is directly obtained from the excitation-damping balance \citep{ivlev18, Chernyshov2024}, 
\begin{equation}
-D\frac{\partial f}{\partial z} \equiv S_D(p,z)= \frac{Bc\nu_{in}}{\pi^2ev_{\rm A}p} \propto\frac{n^{3/2}}{p}  \,,
\label{eq:sdd_definition}
\end{equation}
where $\nu_{in}=\nu_0\, n/n_0$ is the rate of wave damping, scaling linearly with gas density. Thus, 
\begin{equation}
-\frac{\partial}{\partial z}\left(D\frac{\partial f}{\partial z}\right) = \frac{3}{2}\frac{S_{D}}{n}\frac{dn}{dz} \,.
\end{equation}
Since $S_{D}$ does not depend on $B$, here we do not need to make assumptions about $B(z)$. 

Unlike the loss-free case of relativistic CRs studied in \citet{Chernyshov2024}, the flux of lower-energy CRs is generally not conserved, i.e., we need to find a solution of Eq.~(\ref{eq:main_propagation23}). We introduce a new function,
\begin{equation}\tilde{f} = \frac{\dot{p}}{n}v_{\rm A}f \,,
\label{normalization_0}
\end{equation}
and take into account that the ratio $\dot{p}/n\equiv L(p)$ does not depend on density (it is identical to the energy loss function per unit column density; see, e.g., Ref.~\citep{Padovani2009, Padovani2020}). Then, after rearrangement the transport equation can be transformed to
\begin{equation}
\frac{v_{\rm A}}{n} \frac{\partial\tilde{f}}{\partial z} - L\frac{\partial\tilde{f}}{\partial p} = -\frac{3v_{\rm A}LS_{D}}{2n^2}\frac{dn}{dz} \,,
\end{equation}
suggesting the following new variables: 
\begin{equation}
l(z) = \frac{v_{\rm A0}}{n_0}\int \frac{n}{v_{\rm A}} dz\,,\quad \lambda_0(p) = \frac{v_{\rm A0}}{n_0}\int \frac{dp}{L} \,.
\label{variables}
\end{equation}
The choice of the common normalization factor $v_{\rm A0}/n_0\propto n_0^{-3/2}$ for both new variables is such that $l$ would reduce to $z$ in a homogeneous case.\footnote{As in \citet{Chernyshov2024}, we denote the values of parameters at $n = n_i$ by the corresponding indices, e.g., $\left.v_{\rm A}\right|_{n_0} \equiv v_{\rm A0}$, $\left.S_D\right|_{n_1} \equiv S_{D1}$, etc.}

As discussed in \citet{Chernyshov2024}, the density $n_0$ identifies a border through which CRs enter the envelope from the ISM: the border is set at a local minimum of $\nu_{in}$, which is associated with a transition between different dominant ions in the envelope (C$^+$) and the ISM (H$^+$), such that penetrating CRs excite waves only at $n\geq n_0$. 

With the new variables, the transport equation takes the following form: 
\begin{equation}
\frac{\partial\tilde{f}}{\partial l} -  \frac{\partial\tilde{f}}{\partial \lambda_0} = -\frac{3n_0v_{\rm A}LS_{D}}{2v_{\rm A0}n^2}\frac{dn}{dz}\,.
\end{equation}
For $f_0(p)$ being the spectrum of interstellar CRs, the solution is obtained by the method of characteristics,
\begin{equation}
\tilde{f}(\lambda_0,l) = \int\limits_{l_1}^l G(\lambda_0 + l - l', l') dl' + \tilde{f}_0(\lambda_0+l - l_1) \,,
\label{gen_solution}
\end{equation}
assuming the condition $\tilde{f}(\lambda_0,l_1) = \tilde{f}_0(\lambda_0)$ at the outer boundary $l_1=l(z_1)$ (where $n=n_1$ does not necessarily coincides with $n_0$, see Sec.~\ref{first_BC}), and 
\begin{equation}
G(\lambda_0,l) = -\frac{3c \nu_0BL}{2\pi^2ev_{\rm A0}pn}\frac{dn}{dz} \,,
\end{equation}
where the explicit expression for $S_{D}$ is substituted from Eq~(\ref{eq:sdd_definition}), and Eq.~(\ref{variables}) provides the relation to $\lambda_0$ and $l$. Hereafter, $\tilde{f}_0(\lambda_0+l - l_1)$ should be interpreted as
\begin{equation}
    \tilde{f}_0(\lambda_0+l - l_1) \equiv v_{\rm A}L(p_i)f_0(p_i) \,,
    \label{eq:characteristic_explanation}
\end{equation}
where the ``initial'' momentum $p_i$ is derived from
\begin{equation}
    \frac{v_{\rm A0}}{n_0}\int\limits_{p}^{p_i}\frac{dp}{L} =l-l_1\,.
    \label{p_i}
\end{equation}
In analogy to the free-streaming regime \citep{Padovani2018}, Eq.~(\ref{p_i}) defines the ``Alfvenic'' stopping range of trapped CRs moving with the Alfven velocity.

The ionization loss function can be closely approximated by a power-law dependence (see, e.g., Ref.~\citep{Silsbee2019}),  
\begin{equation}
L(p) = {\rm const}~p^{-\beta}\,,   
\label{loss}
\end{equation}
with $\beta \equiv -d\log L/d\log p \approx 1.6$. For protons with the kinetic energy between 0.1~MeV and 0.2~GeV (or for 15~MeV~$\lesssim pc \lesssim$~0.7~GeV, and analogous for heavier nuclei, this deviates by less than 10\% from the exact expression \citep{Padovani2018}. Such energies correspond to the values of stopping range between $\sim10^{19}$~cm$^{-2}$ and $\sim10^{25}$~cm$^{-2}$, so that Eq.~(\ref{loss}) is appropriate to describe continuous losses for all relevant column densities of molecular clouds. We then obtain the loss scale,
\begin{equation}
\lambda_0(p) \approx \frac{v_{\rm A0}}{(1+\beta)n_0}\frac{p}{L} \,, 
\label{lambda_0}
\end{equation}
and 
\begin{equation}
G(\lambda_0,l) \approx -\frac{3c \nu_0B}{2(1+\beta)\pi^2en_0\lambda_0 n}\frac{dn}{dz} \,.
\end{equation}
This allows us to transform Eq.~(\ref{gen_solution}) to the following  solution for the CR spectrum:  
\begin{equation}
f = \frac{\tilde{f}_0(\lambda_0+l - l_1)}{Lv_{\rm A}} -
\frac{3S_{D0}\lambda_0}{2B_0v_{\rm A}}\int\limits_{l_1}^l \left(\frac{B}{n}\frac{dn}{dz}\right)\frac{dl'}{\lambda_0 + l - l'}  \,.
\label{eq:f_solution}
\end{equation}

We assume that $B\approx$~const in diffuse envelopes of molecular clouds \citep{Crutcher2012}, and that the length scale of density inhomogeneity $\Lambda$, defined as
\begin{equation}
    \Lambda \equiv \frac{2n}{3}\left(\frac{d n}{d z}\right)^{-1}\,,
    \label{Lambda_l}
\end{equation}
varies weakly in the envelopes, i.e., $\Lambda \approx$~const. In this case we have $l(n) = \Lambda(n/n_0)^{3/2}\equiv\Lambda(\lambda_0/\lambda)$, which is expressed in terms of the renormalized loss scale,
\begin{equation}
    \lambda(p,n)=\lambda_0(p)\left(\frac{n_0}{n}\right)^{3/2}\,.
    \label{lambda}
\end{equation}
This gives us the following analytic solution for the CR spectrum as a function of density $n$ (replacing coordinate $z$): 
\begin{equation}
f(p,n) \approx  \frac{\tilde{f}_0(\lambda_0\xi)}{Lv_{\rm A}}
-\frac{S_{D0}}{v_{\rm A}} \frac{\lambda_0}{\Lambda} \ln\xi \,,
\label{eq:solution_const_L}
\end{equation}
where for brevity we introduced 
\begin{equation}
    \xi(p,n)\equiv 1 + \Lambda\left(\frac{1}{\lambda}-\frac{1}{\lambda_1}\right)\,,
\end{equation}
and $\lambda_1(p)\equiv\lambda(p,n_1)$. Since the product $S_{D0}\lambda_0$ does not depend on $n_0$, the solution does not depend on it, too.

The behavior of the obtained solution is determined by the ratio $\Lambda/\lambda$: for $\lambda\gg \Lambda$ (the local value of the loss scale is much larger than the inhomogeneity length scale) it is a small parameter, so that Eq.~(\ref{eq:solution_const_L}) can be expanded leading to
\begin{equation}
    f(p,n) \approx f_0(p)\frac{v_{\rm A1}}{v_{\rm A}} - \frac{S_D - S_{D1}}{v_{\rm A}} \,.
\end{equation}
In this case, the CR flux $S(p,n)=S_{D}+v_{\rm A}f(p,n)$ is approximately conserved,
\begin{equation}
    S(p,n) \approx S(p,n_1)\equiv S_1(p) \,,
\end{equation}
i.e., the solution is reduced to that in the loss-free case \citep{Chernyshov2024}. 

\subsection{Fiducial parameters}
\label{fiducial}

For numerical results and examples presented below, we assume the following fiducial values of the model parameters: 

\begin{table}[H]
    \centering
    \begin{tabular}{lll}        
        Length scale of density inhomogeneity:~ & $\Lambda$ & = 5~pc \\
        Magnetic field strength: & $B$ & = 3~$\mu$G  \\
        Mass of the dominant ions: & $m_i$ & = $12m_p$ \\
        Abundance of the dominant ions: & $\xi_i$ & = $1.5\times 10^{-4}$ \\
        Density at the envelope border: & $n_0$ & = 1~cm$^{-3}$ \\
        Wave damping rate at $n=n_0$: & $\nu_0$ & = $9\times 10^{-11}$~s$^{-1}$        
    \end{tabular}
    \label{tab:params}
\end{table}
The value of $\Lambda$ is chosen based on a typical slope of density variations in nearby molecular clouds, as predicted by the 3D dust extinction maps \citep{Edenhofer2024} and illustrated in, e.g., Fig.~1 of \citet{Obolentseva2024}. The selected values for other parameters are discussed in \citet{Chernyshov2024}.

\section{Boundary conditions}
\label{BCs}

Similar to the loss-free case studied in \citet{Chernyshov2024}, there are two conditions regulating the {\it outer} and {\it inner} boundaries of the turbulent zone. We neglect heavier CR nuclei and only study the effect of protons: as was shown in Ref.~\citep{Chernyshov2024} (see their Fig.~2), heavier nuclei significantly contribute to the wave excitation at energies well above $\sim 1$~GeV per nucleon, whereas in the present paper our focus is on the analysis of lower-energy CRs.

\subsection{First boundary condition: outer}
\label{first_BC}

The first condition is set on the outer boundary of the proton diffusion zone, denoted by $n_1(p)$. The condition follows from the requirement that $-D\partial f/\partial z \geq 0$ within the diffusion zone, and therefore $\partial f/\partial z \leq 0$. Using Eq.~(\ref{eq:main_propagation23}), we have for $n>n_0$: 
\begin{equation}
\frac{\partial f}{\partial z} = \frac{n}{v_{\rm A}} \frac{\partial }{\partial p}(Lf) + 
\frac{1}{2}\left(f - \frac{3S_{D}}{v_{\rm A}}\right)
 \left(\frac{1}{n}\frac{dn}{dz}\right) \,.
 \label{df/dz}
\end{equation}
The outer boundary is formed at the ``critical point'' $n=n_{\rm cr}^*$ where $\partial f/\partial z = 0$. Assuming that $\Lambda =$~const and using Eq.~(\ref{lambda_0}) for $\lambda_0$, we reduce this condition to
\begin{equation}
\left.S_{D}\right|_{n_{\rm cr}^*} = \frac{f_{0}\left.v_{\rm A}\right|_{n_{\rm cr}^*}}{3} \left(1 - \frac{3(\alpha_0+\beta)}{1+\beta} \frac{\Lambda}{\left.\lambda\right|_{n_{\rm cr}^*}} \right) \,.
\label{eq:left_condition_losses_pwlw}
\end{equation} 
where 
\begin{equation}
 \alpha_0(p) = -\frac{d\log f_0}{d\log p}    \,,
\end{equation}
is the (momentum-dependent) spectral index of interstellar protons. Substituting expressions for $S_D$ and $v_{\rm A}$, we obtain
\begin{equation}
    \left(\frac{n_{\rm cr}^*}{n_{\rm cr}}\right)^2 = 1- \frac{3(\alpha_0+\beta)}{1+\beta}\frac{\Lambda}{\lambda_0}\left(\frac{n_{\rm cr}^*}{n_0}\right)^{3/2} \,, 
\label{eq:boundary_poly}
\end{equation}
where $n_{\rm cr}(p)$ is the critical point derived by \citet{Chernyshov2024} for the loss-free case of relativistic CRs; its value is given by Eq.~(15) therein:
\begin{equation}
n_{\rm cr}(p) = \sqrt{\frac{\pi pf_0(p)n_0}{12\xi_i} \frac{\Omega_i}{\nu_0}} \,,
\label{eq:n1_noloss}
\end{equation}
where $\Omega_i=eB/m_ic$ is the cyclotron frequency of ions. Equation~(\ref{eq:boundary_poly}) is a fourth-order polynomial for $\sqrt{n_{\rm cr}^*}$. It always has a single positive root $n_{\rm cr}^*\leq n_{\rm cr}$, i.e., $n_{\rm cr}^*/n_{\rm cr}$ decreases in the presence of losses, and the effect is stronger for higher $n_{\rm cr}$.

Following \citet{Chernyshov2024}, we set the outer boundary at $n_1=n_0$ if $n_{\rm cr}^*$ occurs to be smaller than $n_0$. Thus, the outer boundary of the diffusion zone is
\begin{equation}
    n_1(p) = \max\left\{n_{\rm cr}^*(p),\, n_0\right\} \,.
    \label{eq:n1_with_n0}
\end{equation}
The total flux of interstellar CRs entering the diffusion zone is a combination of the diffusion and advection components, $S_1(p)=S_{D1} +v_{\rm A1}f_0$. From Eqs.~(\ref{eq:left_condition_losses_pwlw}) and (\ref{eq:boundary_poly}) we infer
\begin{equation}
\frac{S_{D1}}{v_{\rm A1}f_0}=\frac{1}{3}\left(\frac{n_1(p)}{n_{\rm cr}(p)}\right)^2\,,
\label{K1}
\end{equation}
and hence
\begin{eqnarray}
    S_1(p)=\left[1+\frac{1}{3}\left(\frac{n_1}{n_{\rm cr}}\right)^2\right]v_{\rm A1}f_0\nonumber\hspace{2.5cm}\\
    =\left[1+3\left(\frac{n_{\rm cr}}{n_1}\right)^2\right]S_{D1}\equiv \mathcal{K}S_{D1}\,,
    \label{K2}
\end{eqnarray}
where the factor $\mathcal{K}(p)$ coincides with the expression given by Eq.~(19) of \citet{Chernyshov2024}.

\subsection{Second boundary condition: inner}
\label{second_BC}

The second condition, on the inner boundary $n_2(p)$ of the proton diffusion zone, follows from continuity of their net flux $S$. To derive the condition, let us consider the inner cloud region (beyond the inner boundary) where protons stream freely; we assume that no pre-existing resonant turbulence can be present in such dense regions (see, e.g., Ref.~\citep{Silsbee2019}). The corresponding transport equation is  
\begin{equation}
    \bar{\mu}v\frac{\partial f}{\partial z} - n\frac{\partial}{\partial p}\left(L f\right) + n\sigma v{f} = 0 \,,
    \label{eq:free_propagation_cloud}
\end{equation}
where $v(p)$ is the proton physical velocity and $\sigma$ is the cross section of catastrophic processes associated with the pion production in proton-proton collisions. Isotropic CRs leave the diffusion zone with the mean value of pitch-angle cosine of $\bar{\mu} \approx 1/2$, and for the sake of simplicity we assume this value also for the free-streaming propagation. Below we show that, in case of weak losses in the inner cloud region, the resulting flux does not practically depend on $\bar{\mu}$.

It is convenient to consider a ``reduced'' Eq.~(\ref{eq:free_propagation_cloud}) (without the last term): assuming $\sigma$ to be constant (which is only violated close to the threshold), its solution should then be multiplied by $\exp(-\bar{\mu}^{-1}\mathcal{N}\sigma)$, where
\begin{equation}
    \mathcal{N}(z) = \int n\,dz \,,
\end{equation}
is the gas column density. In order to solve the reduced equation, we introduce a new variable,
\begin{equation}
    R(p) = \int \frac{v}{L}dp  \,,
\end{equation}
which defines the stopping range (column density) in the free-streaming regime, and a new function $\tilde{f} = L(p)f(p,\mathcal{N})$. This leads to equation
\begin{equation}
    \bar{\mu}\frac{\partial \tilde f}{\partial \mathcal{N}} - \frac{\partial \tilde f}{\partial R} = 0 \,.
\end{equation}
For spectrum $f_{\rm in}(p)$ of CRs that stream into the dense clump after leaving the diffusion zone on one side of the cloud, we obtain the solution $f_{\rm out}(p)$ on the other side, 
\begin{equation}
    f_{\rm out}(p,\mathcal{N}) = \frac{\tilde f_{\rm in}(R + \bar{\mu}^{-1}\mathcal{N})}{L} \exp(-\bar{\mu}^{-1}\mathcal{N}\sigma) \,,
    \label{eq:fout_exact}
\end{equation}
where $\tilde f_{\rm in}(R + \bar{\mu}^{-1}\mathcal{N})$ should be interpreted similarly to Eq.~(\ref{eq:characteristic_explanation}). From continuity of the spectrum at the inner boundaries it follows $f_{\rm in} + f_{\rm out} = f_2$, where $f_2(p)$ is given by Eq.~(\ref{eq:solution_const_L}). 

We express the net flux in terms of the derived solution,
\begin{equation}
    S_2(p) = \bar{\mu}v(f_{\rm in} - f_{\rm out}) \equiv uf_2\,,
    \label{S_2}
\end{equation}
where $u(p,\mathcal{N})$ is the flux velocity. For weak losses, one can write $f_{\rm in}\approx \frac12f_2$ and
\begin{eqnarray}
    f_{\rm out} - f_{\rm in} \approx \frac{\mathcal{N}}{\bar{\mu}}\left(\frac{1}{L}\frac{d\tilde{f}_{\rm in}}{d R}-\sigma f_{\rm in}\right)\nonumber\hspace{1.5cm}\\
    =\frac{\mathcal{N}}{\bar{\mu}}\left(\frac{1}{v}\frac{d(Lf_{\rm in})}{d p}-\sigma f_{\rm in}\right)\,.
\end{eqnarray}
Using this result, from Eq.~(\ref{S_2}) we readily obtain $u$ for small $\mathcal{N}$. Given that in the opposite limit of large $\mathcal{N}$ the flux velocity tends to $u\approx \bar{\mu}v$, we interpolate it by
\begin{equation}
    \frac{1}{u} \approx \frac1{\bar{\mu}v} + \frac2{\mathcal{N}}\left((\alpha_2+\beta)\frac{L}{p} + \sigma v\right)^{-1} \,,
    \label{eq:u_approximation}
\end{equation}
where
\begin{equation}
    \alpha_2(p) = -\frac{d\log f_2}{d\log p} \,,
\end{equation}
is the spectral index at the inner boundary.

Now we can write the condition for the inner boundary:
\begin{equation}
    S_2(p)= S_{D2}+v_{\rm A2}f_2 = uf_2 \,.
    \label{eq:n2_differential}
\end{equation}
Since $u$ is a function of $\alpha_2 \propto \partial f_2/\partial p + (\partial f_2/\partial n_2)(dn_2/dp)$, we conclude that Eq.~(\ref{eq:n2_differential}) is a differential equation for $n_2(p)$. The ``initial'' condition for it is set at the excitation threshold $p_{\rm ex}$ -- the maximum momentum at which the proton flux is able to generate waves. The inner and outer boundaries merge at $p=p_{\rm ex}$, so that $f_2 = f_0$ and $\alpha_2 = \alpha_0$. Hence, the value of the threshold is derived from
\begin{equation}
    p_{\rm ex}:\quad S_{D1} + v_{\rm A1}f_0(p_{\rm ex}) = u(p_{\rm ex})f_0(p_{\rm ex}) \,,
    \label{p_ex}
\end{equation}
where $S_{D1}$ and $v_{\rm A1}$ are taken at $n_1(p_{\rm ex})$. Then, $n_2(p_{\rm ex})=n_1(p_{\rm ex})$ yields the sought initial condition for Eq.~(\ref{eq:n2_differential}).

\section{Spectrum and flux of self-modulated protons}

Model spectra of interstellar CRs may be chosen based on available observational data, or they can be approximated by power-law momentum distributions produced in diffusive shock acceleration. The former class is constrained by the AMS-02 measurements \citep{Aguilar2021} with $f_0(p)\propto p^{-2.7}$ for relativistic protons, and by Voyager measurements for non-relativistic protons \citep{Cummings2016, Stone2019}, suggesting $f_0(p)\propto p^{0.2}$ for energies around several MeV. Following \citet{Padovani2018}, one can introduce a broken power-law distribution, 
\begin{equation}
    f_0(p)=\frac{4\pi CE^{-\alpha_0/2}}{(E_{\rm tr}+E)^{2.7-\alpha_0/2}}~{\rm eV}^{-1}~{\rm s}^{-1}~{\rm cm}^{-2}.
    \label{spectrum_L}
\end{equation}
For $C\approx2.4\times10^{15}$, $E_{\rm tr}\approx650$~MeV, and $\alpha_0\approx-0.2$, this corresponds to the proton spectrum measured by the Voyager spacecraft and extrapolated to energies below 3~MeV; it is commonly referred to as ``model $\mathscr{L}$'' \citep{Padovani2018}. The latter class of distributions with a constant spectral index can be presented as
\begin{equation}
    f_0(p)=\frac{4\pi C_{\rm PL} (pc)^{-\alpha_0}}{(2m_pc^2)^{2.7-\alpha_0}}\,,
    \label{spectrum_PL}
\end{equation}
where the normalization is chosen such that $C_{\rm PL}$ and $C$ have the same dimensions; for $C_{\rm PL}=C(2m_pc^2/E_{\rm tr})^{2.7-\alpha_0/2}$, the power-law spectrum (\ref{spectrum_PL}) coincides with the low-energy part of spectrum~(\ref{spectrum_L}).

\begin{figure}
    \centering
    \includegraphics[width=0.9\linewidth]{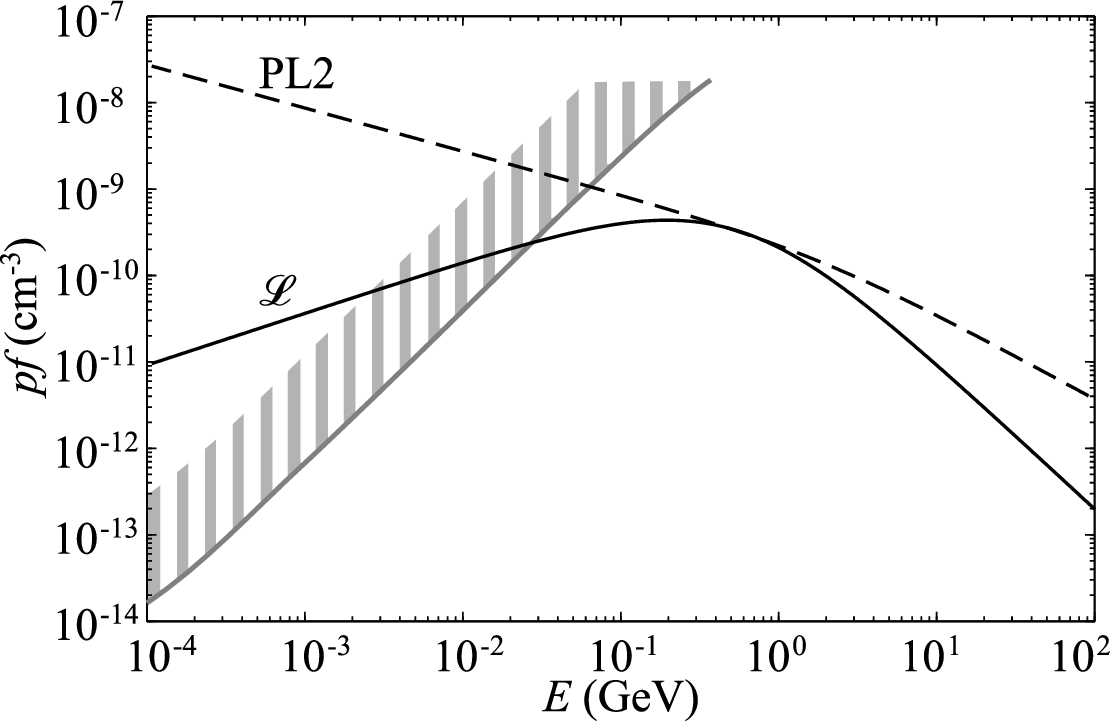}
    \caption{Characteristic spectra of interstellar protons adopted in the present paper: ``model $\mathscr{L}$'' given by Eq.~(\ref{spectrum_L}) with $\alpha_0=-0.2$ (black solid line), and a power-law spectrum given by Eq.~(\ref{spectrum_PL}) with $\alpha_0=2$ (``PL2'', black dashed line). The gray solid line with hatching shows where inequality (\ref{eq:n1_limits}) is violated. 
    }
    \label{fig:n1_limits}
\end{figure}

\begin{figure}
    \centering
    \includegraphics[width=0.9\linewidth]{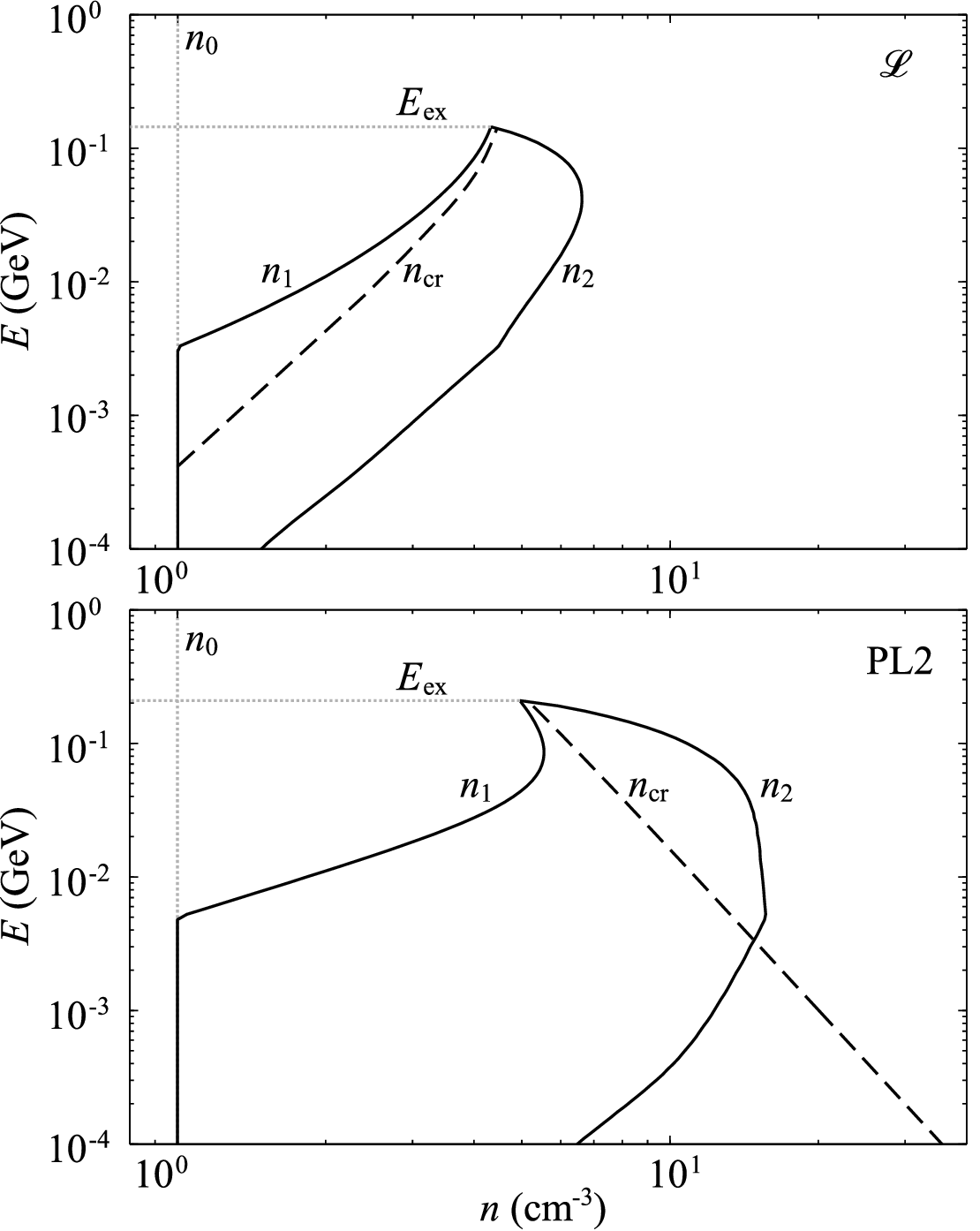}
    \caption{Diffusion zone of CR protons, computed for interstellar spectrum $\mathscr{L}$ (top panel) and PL2 (bottom panel). For clarity, the results are plotted in the plane of proton kinetic energy $E$ (instead of momentum) and gas density $n$. The solid lines show the outer, $n_1(E)$, and inner, $n_2(E)$, boundaries merging at the excitation threshold $E_{\rm ex}$. The dashed lines indicate the (outer boundary) critical density $n_{\rm cr}(E)$ in the loss-free case, given by Eq.~(\ref{eq:n1_noloss}). The results are for the cloud column density of $\mathcal{N} = 10^{22}$~cm$^{-2}$.
    }
    \label{fig:diff_zone}
\end{figure}

\begin{figure*}
    \centering
    \includegraphics[width=0.8\textwidth]{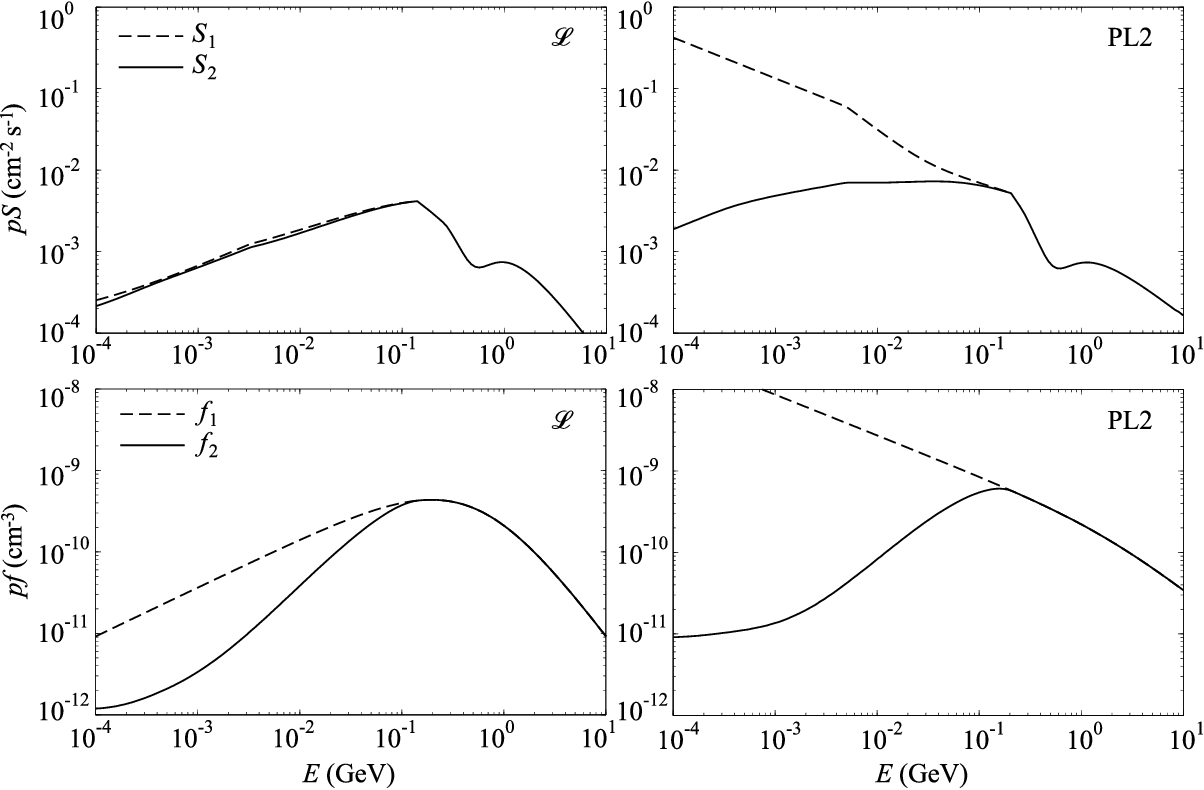}
    \caption{Net flux $S$ (top panels) and spectrum $f$ (bottom panels) of protons at the outer boundary $n_1$ (where $f_1=f_0$, dashed lines) and inner boundary $n_2$ (solid lines) of the diffusion zone. The results are for the interstellar spectrum $\mathscr{L}$ (left) and PL2 (right), assuming the cloud column density of $\mathcal{N} = 10^{22}$~cm$^{-2}$.
    }
    \label{fig:output}
\end{figure*}

Figure~\ref{fig:n1_limits} displays two characteristic realizations of the proton spectra adopted in our paper: model $\mathscr{L}$ and a power-law model with $\alpha_0=2$, hereafter referred to as ``PL2'' (where constant $C_{\rm PL}$ is adjusted such that the two curves touch).  

Figure~\ref{fig:diff_zone} shows the proton diffusion zones computed for the two model spectra. The outer and inner boundaries are derived from Eq.~(\ref{eq:n1_with_n0}) and Eq.~(\ref{eq:n2_differential}), respectively. It is practical to solve Eq.~(\ref{eq:n2_differential}) by employing the following iterative procedure. Knowing $f_2$ (and thus $u$) at a previous iteration step, we compute $n_2$ from Eq.~(\ref{eq:n2_differential}), which allows us to derive $f_2$ for the next iteration from Eq.~(\ref{eq:solution_const_L}). For the first step, we set $f_2 = f_0$. We point out that utilizing interpolation (\ref{eq:u_approximation}) for $u$ may lead to instability of the procedure, and therefore it is preferable to use exact Eqs.~(\ref{eq:fout_exact}) and (\ref{S_2}) for it.

\subsection{Effect of losses}
\label{loss_effect}

Equation~(\ref{eq:boundary_poly}) shows that the critical gas density tends to the loss-free value, $n_{\rm cr}^* \to n_{\rm cr}$, if
\begin{equation}
    \frac{\Lambda}{\lambda_{\rm cr}}\ll \frac{1}{3}\,,
    \label{eq:rel_limit}
\end{equation}
where $\lambda_{\rm cr}\equiv\lambda_0(n_0/n_{\rm cr})^{3/2}$. This inequality can be rewritten in terms of the interstellar spectrum. By substituting Eq.~(\ref{eq:n1_noloss}), we obtain the following condition to neglect losses:
\begin{equation}
    pf_0(p) \ll 
    \frac{4\xi_i n_0\nu_0}{3^{1/3}\pi \Omega_i}\left(\frac{\lambda_0}{\Lambda}\right)^{4/3}\,,
    \label{eq:n1_limits}
\end{equation}
where the rhs scales as $\propto p^{\frac{4}{3}(1+\beta)}$ and does not depend on $n_0$. This condition is shown in Fig.~\ref{fig:n1_limits}: we see the impact is stronger for softer spectra of non-relativistic protons. 

This effect of losses on the outer boundary of the diffusion zone is evident in Fig.~\ref{fig:diff_zone}: it is moderate for a relatively hard spectrum  $\mathscr{L}$, while for soft spectra represented by model PL2 the deviation from the loss-free boundary becomes dramatic. Hence, losses can have a major impact on the diffusion zone, even for relatively high energies. 

Let us now analyze how losses modify the proton flux $S$. Equation~(\ref{K2}) shows that in the regime $n_1=n_{\rm cr}^*(p)$ (corresponding to higher energies in Fig.~\ref{fig:diff_zone}) the proportion between advection and diffusion components exceeds 3/1 in the presence of losses, since $n_{\rm cr}^*/n_{\rm cr}<1$ in this case. At lower energies, where $n_1=n_0$, the proportion depends on the slope of interstellar spectrum: for $\alpha_0<1$ ($>1$), $n_{\rm cr}(p)$ decreases (increases) with decreasing $p$ and, thus, the advection (diffusion) component asymptotically vanishes.  

Using Eq.~(\ref{K2}) along with Eqs.~(\ref{eq:boundary_poly})--(\ref{eq:n1_with_n0}) for the outer boundary and Eqs.~(\ref{eq:n2_differential}) and (\ref{p_ex}) for the inner boundary, we compute the respective fluxes $S_1$ and $S_2$. The modulated spectrum $f_2$ penetrating the cloud is derived from Eq.~(\ref{eq:solution_const_L}). The results for two model interstellar spectra are presented in Fig.~\ref{fig:output}. We see that $S$ is practically conserved for model $\mathscr{L}$, while for PL2 the flux reduction is huge at lower energies. In Appendix~\ref{app1} we present a detailed discussion of this behavior.

We note that in the regimes where the flux remains approximately conserved within the diffusion zone, $S_1\approx S_2$, one can directly evaluate the modulated spectrum from $f_2\approx S_1/u$, using Eq.~(\ref{K2}) with $u$ from Eq.~(\ref{eq:u_approximation}).

Finally, we point out that adiabatic losses can be completely neglected in our problem. In Appendix~\ref{app2} we show that their contribution may only exceed that of regular losses if condition (\ref{eq:rel_limit}) is already satisfied, i.e., adiabatic losses do not affect the diffusion zone at any energy. Similarly, we show that adiabatic losses have no effect on the flux conservation.      

\section{Transport of electrons and their impact on protons}
\label{e_transport}

The resonance between waves having a given wavenumber and CRs occurs for the same value of $p$ for protons and electrons, i.e., the resonant electrons have a much higher velocity than the resonant (non-relativistic) protons. For example, protons with the kinetic energy of 100~keV (and velocity $v_p\approx0.015c$) are resonant with the same waves as ultra-relativistic electrons with the energy of 15~MeV. As a result, such electrons experience much lower energy losses (which are approximately inversely proportional to velocity), and also have a much higher diffusion coefficient (since $D_p/v_p\approx D_e/c$). 

Therefore, if a turbulent zone is generated by CR electrons, one can neglect the diffusion component of proton flux within this zone and assume their advective propagation with the velocity $u\approx v_{\rm A}$. The local Alfven velocity $v_{\rm A}(n)$ decreases toward the cloud center. Hence, if conditions are met for electrons with a given momentum to excite a turbulent zone, then protons with the same momentum may excite their zone only at higher densities, after leaving the electron zone -- where their flux velocity (whose value is set at the inner boundary of the electron zone) exceeds $v_{\rm A}(n)$.

Thus, in regimes where both protons and electrons may generate their own turbulent zones, we need to take into account the following aspects. If the electron turbulent zone is present, the outer boundary of the proton zone may only be located deeper than the inner electron boundary. Therefore, the condition for protons at the outer boundary of their turbulent zone is either set by the advective proton flux leaving the electron zone, or by the interstellar spectrum $f_0$ if no  electron zone is present. The condition at the inner boundary of the proton zone remains unchanged.

For electrons, the condition at the outer boundary of their diffusion zone is $f_e = f_{e0}$, whereas at the inner boundary we need to include backward scattering of electrons by the proton diffusion zone (if present).  

In this paper we assume the interstellar electron spectrum of the form \citep{Padovani2018}
\begin{equation}
    f_{e0}(p)=\frac{4\pi C_e(pc)^{-1.3}}{(E_{{\rm tr},e}+pc)^{1.9}}~{\rm eV}^{-1}~{\rm s}^{-1}~{\rm cm}^{-2},
    \label{spectrum_e}
\end{equation}
where $C_e\approx2.1\times10^{18}$ and $E_{{\rm tr},e}\approx710$~MeV. Similar to the proton model spectrum (\ref{spectrum_L}), Eq.~(\ref{spectrum_e}) provides a fit to the AMS-02 data for $pc\gtrsim1$~GeV, with the slope of $-3.2$ \citep{Aguilar2021}, and describes a transition to the Voyager data for energies down to $\approx3$~MeV, with the slope of $-1.3$ \citep{Stone2019}; the extrapolation to lower energies assumes a constant slope for the momentum distribution.

\subsection{High energies: passive electrons}

For sufficiently high energies electrons are unable to excite turbulence, and thus remain passive CR species. One can also assume that such electrons do not experience substantial losses in the proton diffusion zone.  

The key parameter characterizing the ratio of the advection and diffusion components of the proton flux in the loss-free case is \citep{ivlev18}
\begin{equation}
\begin{split}
    \eta(p) &= \int\limits_{z_1}^{z_2} \frac{v_{\rm A}}{D}dz \equiv -\int\limits_{z_1}^{z_2}\frac{v_{\rm A}}{S_D} \frac{\partial f}{\partial z}dz
\\
    &=
    -\left.\frac{v_{\rm A}}{S_{D}}f\right|_{z_1}^{z_2}
    - 2\int\limits_{n_1}^{n_2} \frac{S}{S_D}\frac{dn}{n} + 2\ln\frac{n_2}{n_1} \,,
\end{split}
\end{equation}
where the last equality is obtained by integrating in parts, keeping in mind that $v_{\rm A}/S_D\propto n^{-2}$, and substituting $v_{\rm A}f = S - S_D$. If $S \approx$~const, which is typical for hard proton spectra (see Appendix~\ref{app1}), the above expression can be further simplified: 
\begin{equation}
\begin{split}
    \eta& \approx \frac{S}{3S_{D2}} - \frac{S}{3S_{D1}} + 2\ln\frac{n_2}{n_1} \\
    &\equiv -\frac{\mathcal{K}}{3}\left[1-\left(\frac{n_1}{n_2}\right)^{3/2}\right]+ 2\ln\frac{n_2}{n_1}\,,
\end{split}    
\end{equation}
which coincides with the loss-free result derived in \citet{Chernyshov2024}, see Eq.~(A3) therein. In our case, the effect of losses is implicitly included here via the value of $n_1$.

The diffusion parameter $\eta_e(p)$ for electrons is different from that for protons by the velocity ratio,\footnote{We use the subscript $e$ to identify parameters attributed to CR electrons, and omit the subscript for protons.}
\begin{equation}
    \eta_e = \frac{v}{v_e}\eta \,.
    \label{eq:eta_e}
\end{equation}
The spectrum of passive electrons at the inner boundary of the proton diffusion zone is then obtained from the (loss-free) solution of Eq.~(\ref{eq:main_propagation23}),
\begin{equation}
    f_e(p,n_2) = e^{\eta_e}\left(f_{e0} - S_{e0} \int\limits_{0}^{\eta_e}\frac{e^{-\eta'_e}}{v_{\rm A} }d\eta'_e\right) \,,
\label{eq:f_integral_form}
\end{equation}
where $S_{e0}(p) = u_e f_e(p,n_2)$ is the electron flux into the cloud determined by the net velocity $u_e(p, \mathcal{N})$ (see Sec.~\ref{secondary_el} below).

Typical values of $\eta$ do not deviate substantially from unity if the  flux of protons is conserved in their diffusion zone. At the same time, velocities of non-relativistic protons are much smaller than velocities of electrons with the same momentum. According to Eq.~(\ref{eq:eta_e}), in this case we have $\eta_e \ll 1$ for electrons with energies well below $\sim1$~GeV. Then, expanding the integral in Eq.~(\ref{eq:f_integral_form}) over $\eta_e$ we readily infer that the electron modulation is negligible if $u_e/v_{\rm A2}\ll v_e/v$. 

A strong electron modulation can be observed for soft spectra of interstellar protons. The role of losses is significant in this case, so that $S\gg S_{D1}$ (see Eq.~(\ref{K2}) and Fig.~\ref{fig:output}), i.e., the advection component of the flux dominates and, hence, $\eta \gg 1$. As a result, values of $\eta_e \sim 1$ can reached, implying a significant modulation. 

\begin{figure}
    \centering
    \includegraphics[width=0.9\linewidth]{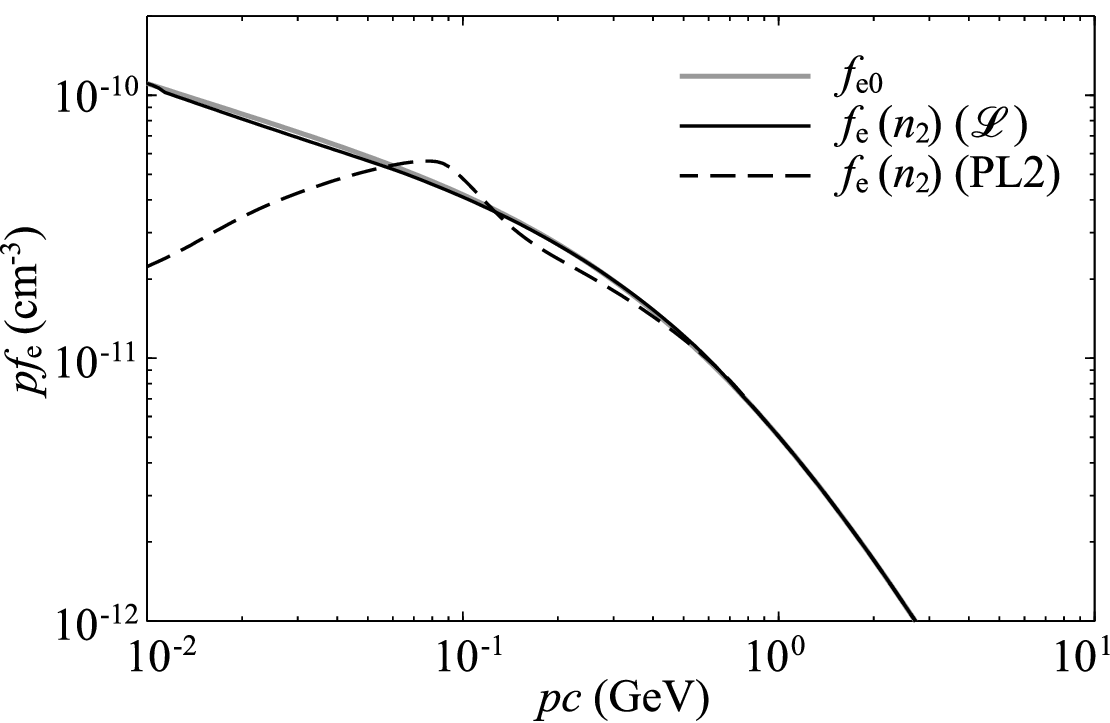}
    \caption{Modulation of CR electrons in the proton diffusion zone, computed from Eq.~(\ref{eq:f_integral_form}). The interstellar electron spectrum, Eq.~(\ref{spectrum_e}), is plotted by the gray line, the modulated spectra at the inner boundary of the diffusion zone are depicted by the black lines, representing  model $\mathscr{L}$ (solid line) and PL2 (dashed line) of interstellar protons. The results are for the cloud column density of $\mathcal{N} = 10^{22}$~cm$^{-2}$.
    }
    \label{fig:electrons}
\end{figure}

Figure~\ref{fig:electrons} illustrates passive modulation of CR electrons in the proton diffusion zone, computed for the two characteristic models of interstellar protons. It is evident that a significant electron modulation is only observed for a soft proton spectrum PL2, whose flux is not conserved. We point out that for sufficiently small values of the flux velocity ($u_e<v_{\rm A}$) it follows from Eq.~(\ref{eq:f_integral_form}) that the spectrum at the inner boundary can exceed the interstellar spectrum.

\subsubsection{Effect of secondary particles}
\label{secondary_el}

The electron flux velocity $u_e(p, \mathcal{N})$ is computed in a similar way to protons, but with a notable difference associated with the effect of secondary electrons and positrons produced in the cloud core. The electron transport equation for the inner cloud region reads 
\begin{equation}
    \bar{\mu}v_e\frac{\partial f_e}{\partial z}-n\frac{\partial}{\partial p}(L_ef_e)=nq_e \,,
    \label{eq:free_propagation_cloud2}
\end{equation}
where $q_e(p,\mathcal{N})$ is the source term of secondary particles produced by protons per hydrogen atom. Unlike Eq.~(\ref{eq:free_propagation_cloud}) for the proton transport, catastrophic (bremsstrahlung) losses in Eq.~(\ref{eq:free_propagation_cloud2}) are included for convenience in the electron loss function $L_e(p)$ \citep{Padovani2018, GinzburgBook1979}.

By introducing the same variables as for Eq.~(\ref{eq:free_propagation_cloud}), we derive the following general solution for $\tilde{f}_e\equiv L_e(p)f_e(p,\mathcal{N})$:
\begin{equation}
    f_{e,{\rm out}}(p,\mathcal{N}) = \frac{\tilde{f}_{e,{\rm in}}(R_e+\bar{\mu}^{-1}\mathcal{N})}{L_e}+f_{e,{\rm sec}}(p,\mathcal{N}) \,,
    \label{f_eout}
\end{equation}
where
\begin{equation}
\begin{split}
    &f_{e,{\rm sec}}(p,\mathcal{N}) \\
    &=\frac{1}{\bar{\mu}L_e}\int\limits_0^\mathcal{N}d\mathcal{N}^\prime~\left.\frac{L_e(p^\prime) q_e(p',\mathcal{N}^\prime)}{v_e(p^\prime)}\right|_{R_e(p^\prime)=R_e(p)+\bar{\mu}^{-1}\mathcal{N^\prime}} \,,
    \end{split}
\end{equation}
and $f_{e,{\rm in}}+f_{e,{\rm out}}=f_e(n_2)$. 

Generally speaking, we need to take into account attenuation of protons for computing $q_e$. However, the problem is substantially simplified for hard proton spectra (such as, e.g., model $\mathscr{L}$), as the generation of secondary particles in this case is dominated by protons with energies over hundreds of MeV. Clouds with column densities of $\mathcal{N} \lesssim 10^{24}$~cm$^{-2}$ are transparent for such protons, and therefore their spectra remain isotropic and equal to $f_2(p)$. Consequently, we can write 
\begin{equation}
   q_e(p) \approx \frac{1}{2}\int\limits_{p'_{\rm min}}^\infty v(p')f_{2}(p')\frac{\partial\sigma}{\partial p}(p',p)\, dp' \,,
\end{equation}
where the integration is over the proton momentum. The differential cross section of proton-impact collisions, $\partial\sigma/ \partial p$, includes both the ionization and generation of secondary electrons and positrons upon decay of the charged pions, with $p'_{\rm min}$ being the minimum proton momentum needed to eject an electron with momentum $p$. We note that the result obtained in this approximation does not depend on a particular gas distribution in the cloud. 

Assuming weak electron losses in the core, i.e., $f_{e,{\rm in}}\approx f_{e,{\rm out}}$ we then derive $u_e(p,\mathcal{N})$ following the approach described in Sec.~\ref{second_BC}.

\subsection{Onset of electron-driven turbulence}
\label{el_turbulence}

Knowing how CR electrons propagate in the passive regime, we now can determine the electron excitation threshold $p_{{\rm ex},e}$ -- the maximum momentum at which the electron flux is able to excite turbulence.
 
Consider the regime with $p<p_{{\rm ex},e}$. Irrespective of whether the proton diffusion zone exists (see discussion in the beginning of Sec.~\ref{e_transport}), the outer boundary of the electron diffusion zone, $n_{1,e}(p)$, is obtained by applying Eq.~(\ref{eq:n1_noloss}) for the interstellar electron spectrum $f_{e0}(p)$. The electron flux formed at that boundary,\footnote{To distinguish between parameters at the boundaries of proton and electron diffusion zones, their values at the electron zone boundary are denoted by, e.g., $S_{D1,e}$ or $v_{{\rm A1},e}$ at the boundary $n_{1,e}$.} 
\begin{equation}
S_{e0}(p) = S_{D1,e}+ v_{{\rm A1},e}f_{e0}\,,     
\end{equation}
is then conserved. 
 
For solving a problem with coexisting electron and proton diffusion zones, it is more convenient to directly compute the electron spectrum at the inner boundary of the proton zone, instead of deriving the value of $u_e$. As we consider relativistic electrons and assume moderate column densities, so that $f_{e,{\rm in}}\approx f_{e,{\rm out}}=\frac12f_e|_{n_2}$, one can integrate Eq.~(\ref{eq:free_propagation_cloud2}) over $z$ across the cloud and over $p$ to infinity.\footnote{Electron spectra at the proton zone boundaries are denoted by, e.g., $f_{e}|_{n_1}$ at the boundary $n_1$; similarly, proton spectra (and also fluxes) at the electron zone boundaries are, e.g., $f|_{n_{1,e}}$ at the boundary $n_{1,e}$.} Using the relation $ S_{e0} = \bar{\mu}v_e(f_{e,{\rm in}} - f_{e,{\rm out}})$, this allows us to write the electron spectrum at the inner boundary of the proton diffusion zone in the following form:
\begin{equation}
    f_{e}|_{n_2} = \frac{2}{\mathcal{N}L_e}\int\limits_p^\infty S_{e0}\,dp'  + \frac{2}{L_e}\int\limits_p^\infty q_e\,dp' \,.
    \label{integral}
\end{equation}
It is worth noting that the momentum of protons generating secondary particles is always much higher than the resulting momentum of those particles. Therefore, the electron diffusion zone has no effect on the spectrum of such protons and thus on $q_e(p)$.

If the problem of proton propagation is solved and the value of $\eta_e$ is known, we can now utilize Eq.~(\ref{eq:f_integral_form}) to infer the electron spectrum at the outer boundary of the proton diffusion zone,  
\begin{equation}
    f_{e}|_{n_1} = e^{-\eta_e}f_e|_{n_2} + S_{e0} \int\limits_{0}^{\eta_e}\frac{e^{-\eta'_e}}{v_{\rm A} }d\eta'_e \,.
\end{equation}
This spectrum should be formed in the electron diffusion zone. Hence, the inner boundary of this zone, $n_{2,e}(p)$, is derived from the following matching condition:
\begin{equation}
    f_{e}|_{n_1}=f_{e2}\;,
\end{equation}
where the electron spectrum $f_{e2}(p)$ at the boundary is obtained from the relation $S_{e0}(p) =S_{D2,e}+ v_{{\rm A2},e}f_{e2}$. The electron diffusion zone exists if $n_{2,e}\geq n_{1,e}$, and therefore the electron excitation threshold is given by the following condition:
\begin{equation}
p_{{\rm ex},e}:\quad n_{2,e}(p_{{\rm ex},e}) = n_{1,e}(p_{{\rm ex},e})\,.
\end{equation}

\begin{figure}
    \centering
    \includegraphics[width=0.95\linewidth]{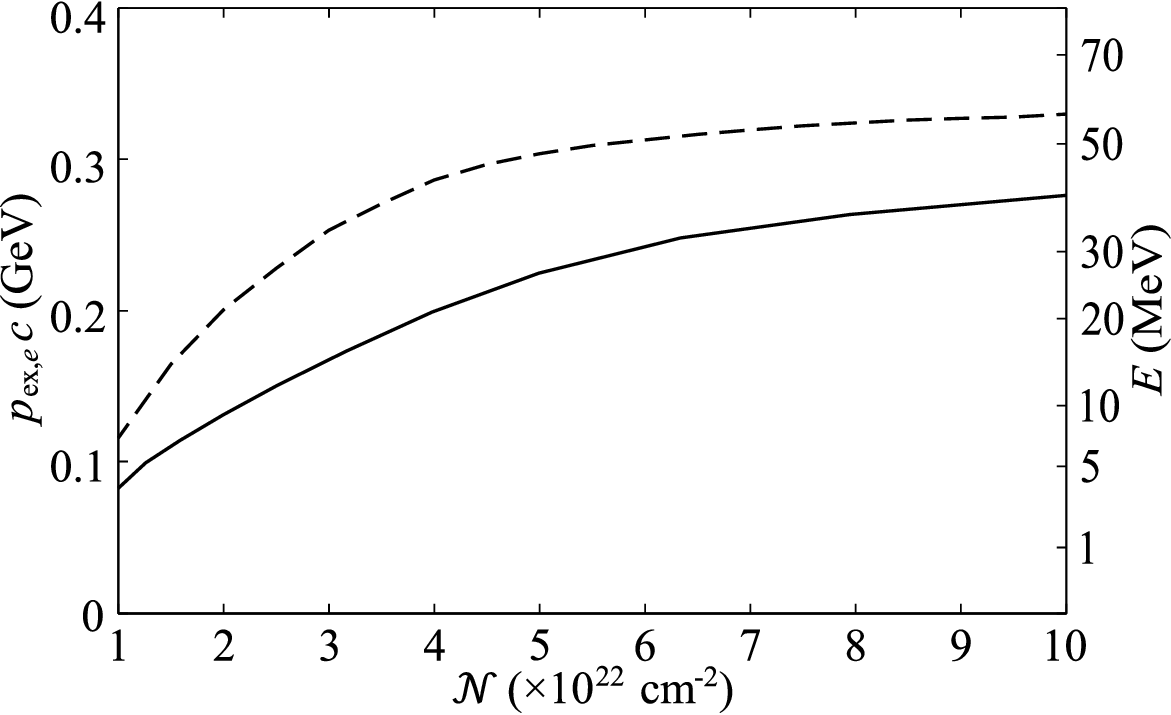}
    \caption{Electron excitation threshold, $p_{{\rm ex},e}$, derived for the interstellar spectrum (\ref{spectrum_e}) and plotted versus the cloud column density $\mathcal{N}$. To demonstrate the effect of secondary particles, the solid and dashed lines show the results computed, respectively, with and without taking into account the source term in Eq.~(\ref{integral}). The right axis displays the corresponding resonant energy of protons.
    }
    \label{fig:electrons_pex}
\end{figure}

Figure~\ref{fig:electrons_pex} shows the electron excitation threshold  plotted versus the cloud column density. We consider two cases: with and without generation of secondary particles. In the latter case the net electron flux is naturally decreased and, therefore, the value of $p_{{\rm ex},e}$ is reduced. For reference, the right axis displays the corresponding resonant energy of protons.

\subsection{Low energies: passive protons}

Given that the ratio of proton and electron diffusion coefficients is equal to their velocity ratio, the diffusion component of the proton flux in the electron zone can be reasonably neglected, i.e., we can assume $S\approx v_{\rm A}f$ within the zone. Then, in order to describe propagation of passive protons with $p<p_{{\rm ex},e}$, we only need to know boundaries of the electron diffusion zone, $n_{1,e}(p)$ and $n_{2,e}(p)$.

The proton flux in the advection-dominated regime can be directly deduced from the second term of Eq.~(\ref{gen_solution}). The flux leaving the electron diffusion zone is 
\begin{equation}
    S|_{n_{2,e}} = v_{{\rm A1},e}\frac{\tilde{f}_0(\lambda_0+l_{2,e}-l_{1,e})}{L} \,,
    \label{eq:protons_passive}
\end{equation}
(here $\tilde{f}_0=Lf_0$). Equation~(\ref{eq:u_approximation}) suggests that the velocity of proton flux formed in a sufficiently dense cloud could exceed the Alfven velocity in the cloud envelope, which would lead to the excitation of an additional diffusion zone. In this case, the problem is reduced to that discussed in  Sec.~\ref{BCs}, where we only need to replace $f_0$ at the outer boundary $n_1$ of the proton zone with 
\begin{equation}
    f_1 = \frac{S|_{n_{2,e}}-S_{D1}}{v_{\rm A1}} \,.
\end{equation}
Substituting this in Eq.~(\ref{df/dz}), we derive $n_1$ from the condition $\partial f/\partial z|_{n_1} = 0$. Obviously, the necessary condition for the proton diffusion zone to form in this case is that the resulting $f_1$ is positive. If the proton diffusion zone does not form, the modulated spectrum of protons is simply $f|_{n_{2,e}} =S|_{n_{2,e}}/u$, where $u(p,\mathcal{N})$ is given by Eq.~(\ref{eq:u_approximation}). 

The outer boundary of the proton zone is detached from the electron zone if $n_1>n_{2,e}$, otherwise it starts immediately at the inner boundary of the electron zone. The inner proton boundary $n_2(p)$ is derived from the same boundary condition as in the regular case described by Eq.~(\ref{eq:n2_differential}).

In order to self-consistently include mutual effects of coexisting proton and electron diffusion zones, we use an iterative procedure. We start with the proton zone boundaries computed without taking into account the possible existence of the electron zone -- whose boundaries are derived as described in Sec.~\ref{el_turbulence}. Then we recompute the proton zone boundaries following the procedure discussed above in this section. If the proton diffusion zone is formed, we repeat the iterative procedure until it converges; if not, we re-derive the electron zone boundaries without the effect of the proton zone.

\begin{figure*}
    \centering
    \includegraphics[width=0.85\textwidth]{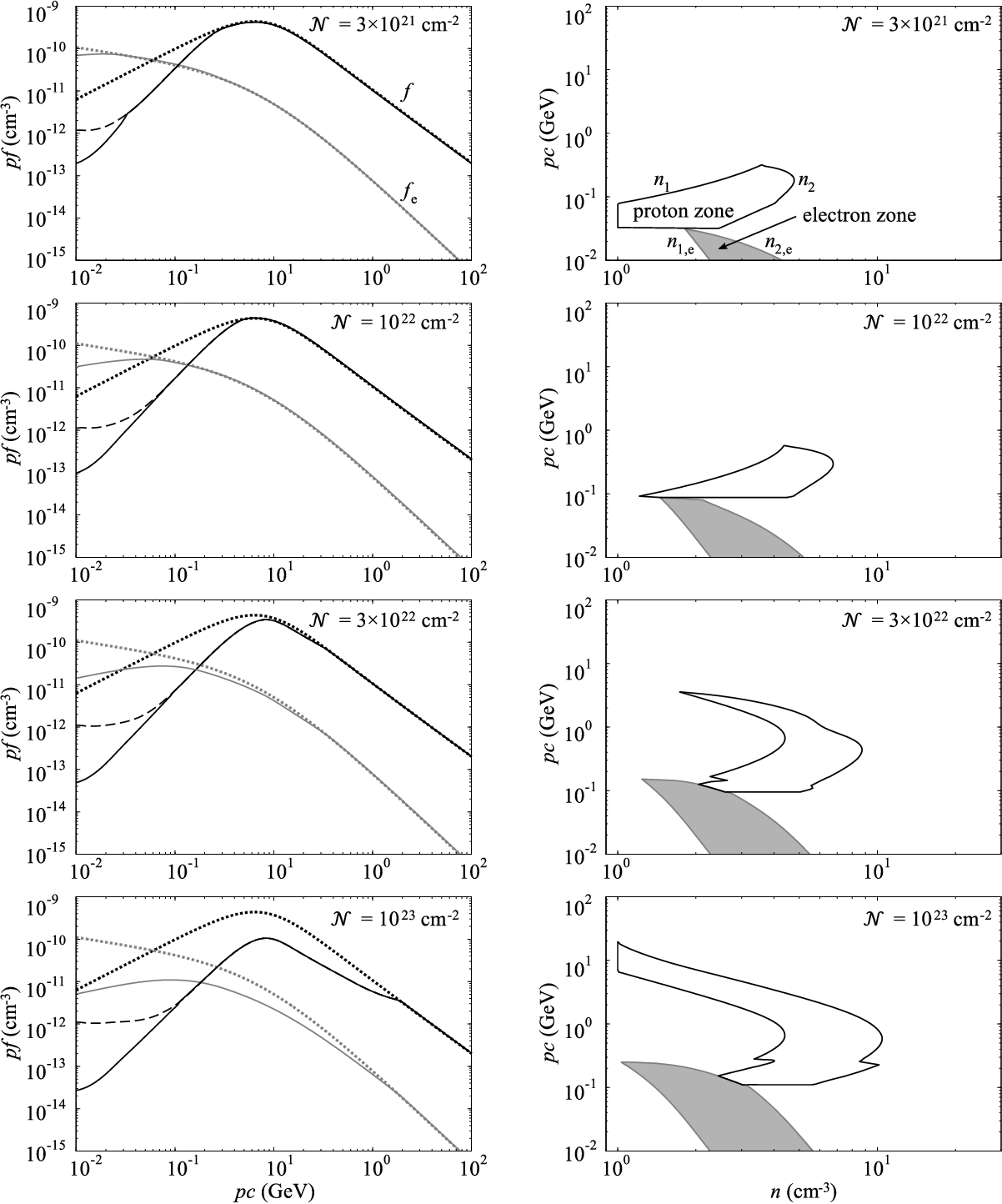}
    \caption{CR spectra (left panels) and diffusion zones (right panels) for different values of the cloud column density $\mathcal{N}$. Proton and electron spectra ($f$ and $f_e$) are depicted by the black and gray lines, respectively. The solid lines show the modulated spectra, computed at the inner(most) boundary of the diffusion zone, the dotted lines are the interstellar spectra, given by Eq.~(\ref{spectrum_L}) with $\alpha_0=-0.2$ (model $\mathscr{L}$) for protons and by Eq.~(\ref{spectrum_e}) for electrons. For comparison, the black dashed lines show the modulated proton spectra derived neglecting the electron-driven turbulence. In the right panels, the proton and electron diffusion zones are depicted by the unshaded and shaded contours, respectively.
    }
    \label{fig:composite}
\end{figure*}

The modulated spectra of CR protons and electrons as well as the corresponding diffusion zones, computed for model $\mathscr{L}$ of interstellar protons, are presented in Fig.~\ref{fig:composite} for different $\mathcal{N}$. For lower values of $\mathcal{N}$ the proton zone is ``replaced'' by the electron zone at $p<p_{{\rm ex},e}$, whereas for higher $\mathcal{N}$ the two zones coexist within a certain momentum range.

We point out that for sufficiently soft spectra of interstellar protons, such as model PL2, the values of $p_{{\rm ex},e}$ are very small, i.e., CR electrons remain passive in the considered range of energies.

\subsection{Transition from Alfven waves to whistlers}

The model of electron diffusion zone discussed above is applicable for energies where excited turbulence is Alfvenic. This assumption is valid as long as the Alfven frequency $v_{\rm A}k$ is smaller than the cyclotron frequency of carbon ions $\Omega_i$. Utilizing the resonant relation $k=eB/pc$ we obtain the following condition:
\begin{equation}
    pc \gg m_iv_{\rm A}c\approx 5\times10^{-3}\left(\frac{B}{3~\mu\mbox{G}}\right)\left(\frac{n}{1~\mbox{cm}^{-3}}\right)^{-1/2}~\mbox{GeV}.
    \label{eq:p_whistler}
\end{equation}
For smaller $p$ we need to keep in mind that the resonant waves change from the Alfven to the whistler branch, whose dispersion relation is given by (see, e.g., Ref.~\citep{GinzburgBook1970})
\begin{equation}
    \omega\approx\frac{\Omega_e+i\nu_e}{\omega_{pe}^2}c^2k^2\,,
\end{equation}
where $\Omega_e=eB/m_ec$ is the electron cyclotron frequency, $\omega_{pe}$ is the electron plasma frequency, and $\nu_e$ is the momentum transfer rate for electron collisions (which generally includes collisions with neutrals and ions). Using the identity $\Omega_e\Omega_ic^2\equiv \omega_{pe}^2v_{\rm A}^2$, we derive the phase velocity of whistlers,
\begin{equation}
    v_{\rm w}=\frac{v_{\rm A}^2k}{\Omega_i} \equiv v_{\rm A}^2\frac{m_i}{p} \,,
\end{equation}
and their damping rate,
\begin{equation}
    \nu_{\rm w} = \nu_e\frac{m_em_iv_{\rm A}^2}{p^2} \equiv\nu_e\frac{ m_ev_{\rm w}}{p} \,.
\end{equation}

The rate of resonant wave excitation by streaming CRs is proportional to the phase velocity of the waves \citep{KulsrudBook2005}, i.e., in the regular expression for the excitation rate \citep{Skilling1975, Chernyshov2024} we now need to replace $v_{\rm A}$ with $u_{\rm w}$,
\begin{equation}
    \gamma_{\rm w}\approx-\frac{\pi^2ev_{\rm w}p}{m_ec^2\Omega_e}D\frac{\partial f}{\partial p} \,.
\end{equation}
Hence, the excitation-damping balance reads
\begin{equation}
\gamma_{\rm w}=\nu_{\rm w} \,.
\label{Wave_Eq}
\end{equation}
We note that both terms in Eq.~(\ref{Wave_Eq}) are proportional to $u_{\rm w}$ and thus the latter cancels out. However, one can see that the relative magnitude of the damping rate for whistlers is changed, compared to that for Alfven waves, by a factor of $(\nu_e/\nu_{in})(m_ev_{\rm A}/p)$. Given that the ratio $v_{\rm A}/v_e$ is very small for relevant electron energies, we conclude that the relative damping rate is reduced in the regime of whistler excitation.

\begin{figure}
    \centering
    \includegraphics[width=0.9\linewidth]{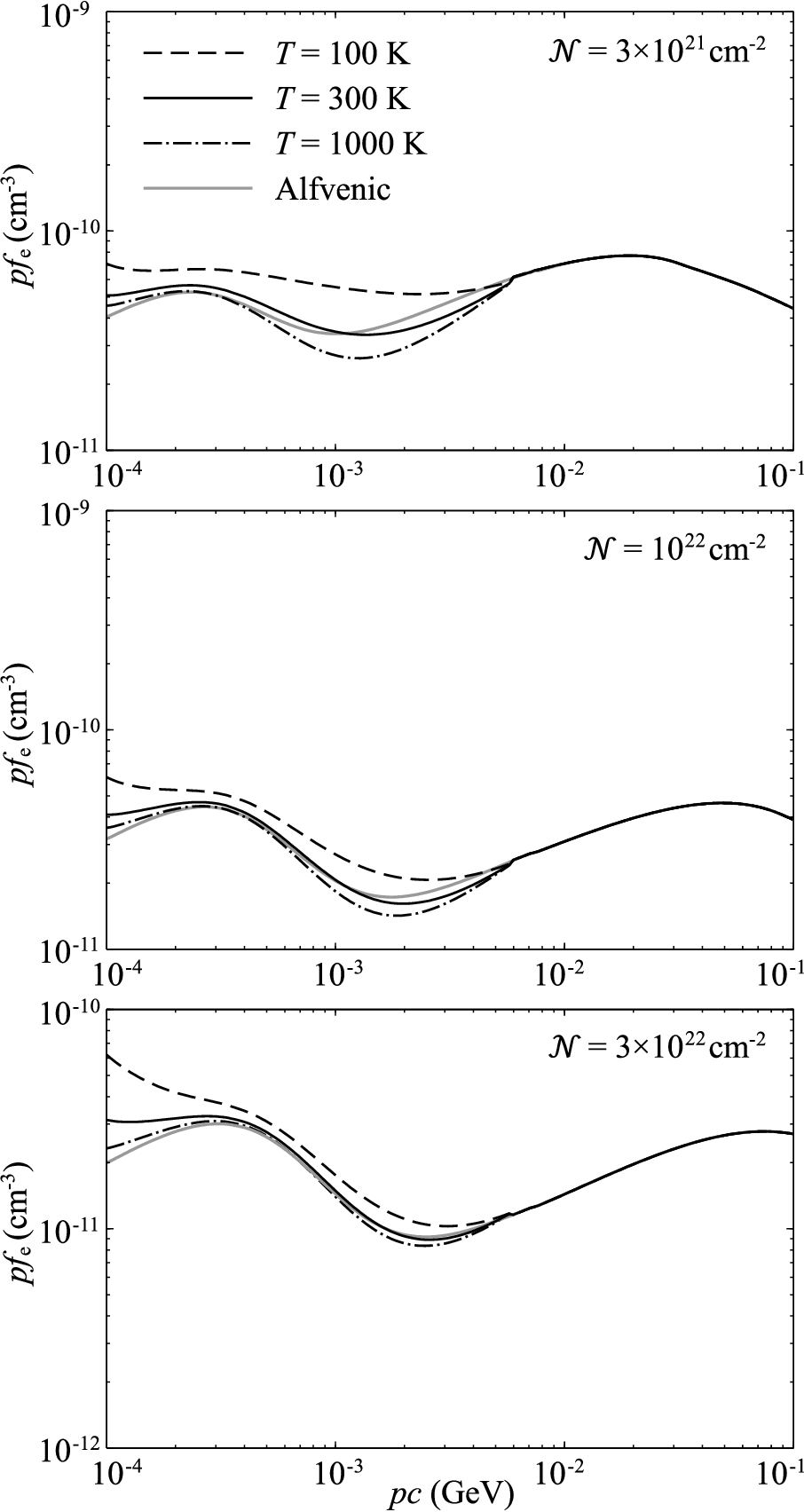}
    \caption{Spectrum of modulated CR electrons at the inner boundary of the electron diffusion zone, derived for the interstellar spectrum (\ref{spectrum_e}) and plotted for different values of the gas temperature $T$. Transition between Alfvenic and whistler regimes occurs at $pc = 6\times 10^{-3}$~GeV. Different panels show the results for different values of the cloud column density $\mathcal{N}$.
    }
    \label{fig:electrons_full}
\end{figure}

\begin{figure}
    \centering
    \includegraphics[width=0.9\linewidth]{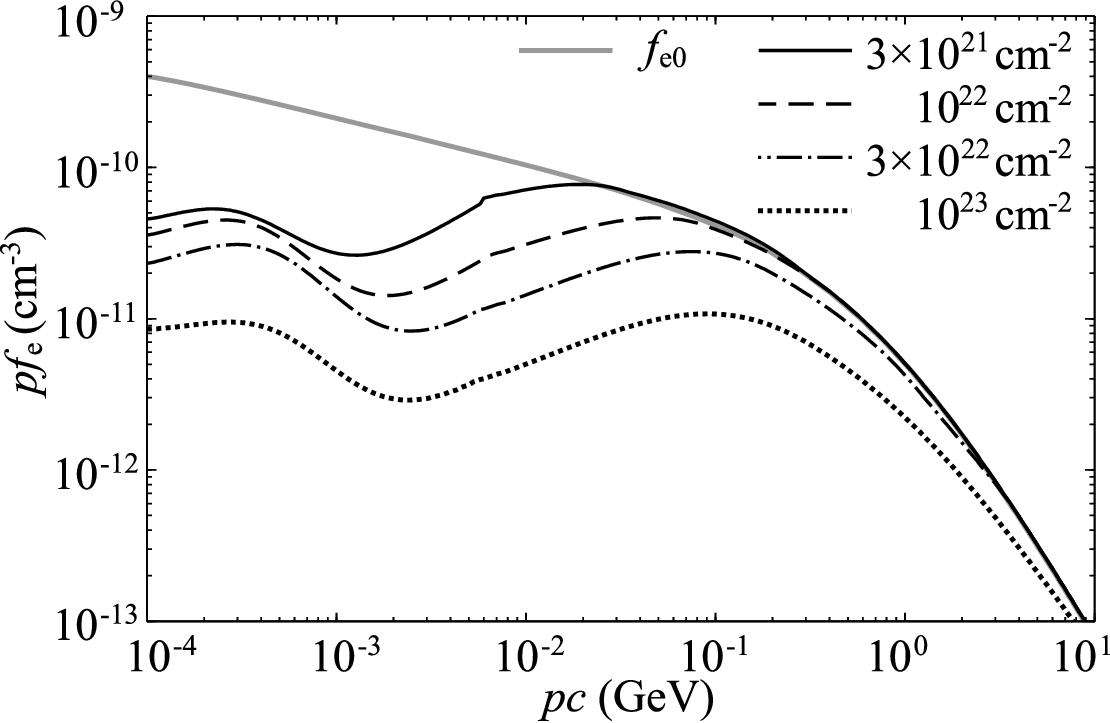}
    \caption{Modulated electron spectrum for molecular clouds with different column densities (see Fig.~\ref{fig:electrons_full}), assuming the gas temperature in the envelopes of $T=1000$~K.
    }
    \label{fig:electrons_full2}
\end{figure}

Equation~(\ref{Wave_Eq}) yields the diffusion flux in the whistler regime,
\begin{equation}
    S_{D{\rm w}} = \frac{Bc\nu_em_e}{\pi^2ep^2}\propto \frac{n}{p^2} \,.
\end{equation}
Comparing this to Eq.~(\ref{eq:sdd_definition}), we see a different scaling dependence and also a different magnitude of the flux: the Alfven velocity in the denominator of $S_D$ is now replaced with a much larger value of $p/m_e$. 

The electron transport equation in the whistler regime is 
\begin{equation}
\frac{\partial}{\partial z}\left( v_{\rm w}f_e +S_{D{\rm w}}\right) - n\frac{\partial}{\partial p}\left(L_ef_e\right) = 0\,.
\end{equation}
Since $v_{\rm w}\propto (np)^{-1}$, we need to introduce new variables that are different from those used to solve the problem in Sec.~\ref{gov_eqs}. Now, for $\tilde{f}_e = L_ev_{\rm A}^2f_e$ we use
\begin{equation}
    l_{\rm w}(z)=\frac{v_{\rm A0}^2}{n_0}\int\frac{n}{v_{\rm A}^2}dz\,,\quad\lambda_{\rm w}(p)=\frac{m_iv_{\rm A0}^2}{n_0}\int\frac{dp}{pL_e} \,,
\end{equation}
which allows us to reduce the equation to
\begin{equation}    
\frac{\partial \tilde{f}_e}{\partial l_{\rm w}}-\frac{\partial \tilde{f}_e}{\partial \lambda_{\rm w}}=-\frac{2pL_eS_{D{\rm w}}}{3m_i\Lambda}\left(\frac{n_0}{n}\right)^2 \,.
    \label{eq:e_fe_whistlers}
\end{equation}
Its general solution is then obtained from Eq.~(\ref{gen_solution}).

The outer boundary of the diffusion zone, $n_{1,e}(p)$, is derived from the following equation similar to Eq.~(\ref{df/dz}):
\begin{equation}
    \frac{\partial f_e}{\partial z}= \frac{n}{v_{\rm w}}\frac{\partial }{\partial p}(L_ef_{e}) + \frac{2}{3\Lambda}\left(f_{e}-\frac{S_{D{\rm w}}}{v_{\rm w}}\right)\,,
\end{equation}
with the conditions $\partial f_e/\partial z = 0$ and $f_e=f_{e0}$. The condition at the inner boundary, $n_{2,e}$, also has a similar form:
\begin{equation}
    S_{D{\rm w}2,e}+ v_{{\rm w2},e}f_{e0}   = \bar{\mu}v_e(f_{e,{\rm in}} - f_{e,{\rm out}}) \,,
    \label{eq:inner_boundary_e_whistler}
\end{equation}
where the rhs is computed from Eq.~(\ref{f_eout}). The resulting velocity of the net electron flux is always small, $u_e \ll v_e$, due to the effect of secondary particles, as discussed in Sec.~\ref{secondary_el}. 

The computed electron spectra are plotted in Fig.~\ref{fig:electrons_full} for different values of gas temperature $T$ in diffuse envelopes of molecular clouds and for different cloud column densities $\mathcal{N}$. We solve Eq.~(\ref{eq:e_fe_whistlers}) for $pc < 6\times 10^{-3}$~GeV, and use the Alfvenic solution (\ref{eq:solution_const_L}) for higher $p$; for comparison, we also present the results obtained from Eq.~(\ref{eq:solution_const_L}) only. The growth seen in the spectra toward smaller $p$ is due to increasing contribution of secondary particles. Note that $pc=10^{-4}$~GeV correspond to non-relativistic electrons with the kinetic energy $\approx(pc)^2/ 2m_ec^2 \sim 10$~keV.

The gas temperature enters the results via the electron collision rate $\nu_e$, which is dominated by the Coulomb collisions with ions for all temperatures expected in the WNM and CNM \citep{DraineBook2011}, scaling as $\nu_e\propto T^{-3/2}$. However, Fig.~\ref{fig:electrons_full} shows that for $T\gtrsim$~300~K (relevant for diffuse envelopes) the derived electron spectra are practically insensitive to the temperature value. The reason behind is that at such temperatures $S_{D{\rm w}2,e}\propto \nu_e$ rapidly decreases and becomes smaller than the contribution of secondary electrons in the rhs of Eq.~(\ref{eq:inner_boundary_e_whistler}) [see Eq.~(\ref{f_eout})]. In Fig.~\ref{fig:electrons_full2} we illustrate dependence of the modulated electron spectrum on the gas column density, assuming $T=1000$~K.

\section{Conclusions}
\label{conclusions}

The theory of CR penetration into dense molecular clouds developed for relativistic particles by \citet{Chernyshov2024} has been extended to non-relativistic CRs. We showed that, while such CRs undergo significant ionization losses in diffuse envelopes surrounding the clouds, the mechanism of self-modulation operating in nonuniform envelopes remains essentially unchanged. At the same time, the self-modulation of non-relativistic CRs is much more efficient, and can be substantial even for clouds with moderate column density of $\mathcal{N}\gtrsim10^{21}$ cm$^{-2}$. Our main finding can be summarized as follows:
\begin{enumerate}[(1)]
    \item Boundaries of a turbulent zone depend on the dimensionless ratio of the density inhomogeneity length scale in the envelope $\Lambda$ to the local loss scale $\lambda_{\rm cr}$ [Eq.~(\ref{eq:rel_limit})]: for large $\Lambda/\lambda_{\rm cr}$ losses dramatically modify the outer boundary $n_1(p)$ [Eqs.~(\ref{eq:boundary_poly}) and (\ref{eq:n1_with_n0})], shifting it to a lower density, while in the opposite limit the outer boundary tends to the loss-free value [Eqs.~(\ref{eq:n1_noloss}) and (\ref{eq:n1_with_n0})] derived in \citet{Chernyshov2024}. The effect depends on the interstellar spectrum of CRs and is stronger for softer spectra, as illustrated in Fig.~\ref{fig:diff_zone}.  
    \item Even if $\Lambda/\lambda_{\rm cr}$ is large and a turbulent zone is strongly modified by losses, the net flux of penetrating CRs is practically conserved for a fairly broad class of sufficiently hard interstellar spectra, such as, e.g., the proton spectrum measured by the Voyager spacecraft (see Fig.~\ref{fig:output}). The modulated spectrum in this case can be computed using formulas derived in \citet{Chernyshov2024} for the loss-free case (but taking into account the shift of the outer boundary).
    \item A turbulent zone can be excited both by CR protons and electrons. Protons generally dominate at higher momentum values, where electrons play a role of a passive component modulated by proton-generated waves. As soon as the threshold condition for electron-driven turbulence is met at a lower momentum, protons become a passive component. For sufficiently high $\mathcal{N}$, a regime where both protons and electrons generate their own turbulent zones is realized within a certain momentum range; in this case, the outer boundary of the proton zone is always located deeper than the inner boundary of the electron zone, as depicted in Fig.~\ref{fig:composite}. 
    \item At lower momentum values, where the turbulent zone is generated by electrons only, their net inward flux becomes gradually limited due to production of secondary electrons and positrons in the cloud interior, which correspondingly reduces the modulation depth (see Fig.~\ref{fig:electrons_full2}).
\end{enumerate}

We conclude that the self-modulation can be considered as an efficient mechanism of  ``unification'' of CR spectra penetrating molecular clouds, making (potentially) diverse interstellar spectra similarly hard. This phenomenon is illustrated in Fig.~\ref{fig:output}, showing that very different spectra assumed for interstellar CRs become much more alike inside the clouds. Given that turbulent zones are generated in outer regions of diffuse envelopes, corresponding to gas densities of $\sim1-10$~cm$^{-3}$, we expect that our results may be relevant for understanding nearly all available measurements of the CR ionization rate in molecular gas. Furthermore, combining the results of the present paper with our earlier results reported in \citet{Chernyshov2024} may in future help us to formulate a consistent view on the impact of Galactic CRs on ionization and gamma-ray emission produced in molecular clouds \citep{Padovani2020, Gabici2022,Phan2020,Fujita2021,Becker2011, Vaupre2014, Owen2021, dogiel2015, Ravikularaman2025, Indriolo2013}.

\begin{appendix}

\section{Variation of flux across the diffusion zone}
\label{app1}

Let us assume that $\Lambda \approx$~const in the diffusion zone, and try to evaluate conditions where $S(z) \approx$~const. For this purpose, we substitute $f \approx (S - S_D)/v_{\rm A}$ in the loss term of Eq.~(\ref{eq:main_propagation23}) and integrate the result over $z$ between $z_1$ and $z_2$. We obtain
\begin{eqnarray}
    S_2 - S_1 = \frac{\Lambda n_1}{v_{\rm A_1}}\left[\left(\frac{n_2}{n_1}\right)^{3/2} - 1\right]\frac{\partial(LS)}{\partial p}\nonumber\hspace{1.5cm}\\ 
    -\frac{\Lambda n_1}{2v_{\rm A_1}}\left[\left(\frac{n_2}{n_1}\right)^{3} - 1\right]\left.\frac{\partial(LS_D)}{\partial p}\right|_{n_1(p)} \,,
\end{eqnarray}
where $\partial(LS_D)/\partial p=-(1+\beta)LS_D/p$. On the other hand, $S(p) \approx S_{D1}+ v_{\rm A1}f_0$ depends on $p$  explicitly, via $f_0(p)$ and $S_{D1}(p)$, and also implicitly, via $n_1(p)$ if $n_1>n_0$. However, $n_1$ is the critical point for $S_0$, i.e., $\partial S/\partial n|_{n_1} = 0$ if losses do not play a substantial role. Hence,
\begin{equation}
    \frac{\partial(LS)}{\partial p} \approx - (1+\beta)\frac{LS_D}{p} - (\alpha_0+\beta) \frac{Lv_{\rm A}f_0}{p} \,.
\end{equation}
Introducing $X \equiv S_{D1}/(v_{\rm A1}f_0)$ for brevity and utilizing Eqs.~(\ref{K1}) and (\ref{K2}) as well as Eq.~(22) from \citet{Chernyshov2024}, one can estimate the density ratio as $(n_2/n_1)^{3/2} \approx 1 + X^{-1}$. We finally arrive to
\begin{equation}
    \frac{S_1 - S_2}{S} = \frac{\Lambda}{\lambda_1} \frac{\alpha_0 + \beta/2 - 1/2}{X(1+\beta)(1 + X)} \,,
    \label{eq:flux_conservation}
\end{equation}
where $X=\frac13(n_1/n_{\rm cr})^2$ according to Eq.~(\ref{K1}). We see that the flux is conserved if $\alpha_0 = (1-\beta)/2\approx-0.3$. Remarkably, this value is quite different from $\alpha_0 = -1.6$ which would follow from a steady-state solution $f(p)\propto L^{-1}$. 

We see that the relative flux variation is determined by the magnitude of the factor $(\Lambda/\lambda_1)/X$. From the positiveness of the rhs of Eq.~(\ref{eq:boundary_poly}) we readily derive that 
\begin{equation}
    \frac{\Lambda}{\lambda_1}\frac{1}{X}\lesssim 
\frac{(n_{\rm cr}/n_1)^{1/2}}{(n_{\rm cr}^*/n_{\rm cr})^{3/2}}\,.
    \label{factor}
\end{equation}
We note that the rhs of Eq.~(\ref{factor}) is reduced to $(n_{\rm cr}^*/n_{\rm cr})^{-2}$ in the regime $n_1=n_{\rm cr}^*\;(>n_0)$, which is precisely the inverse of the ratio in Eq.~(\ref{eq:boundary_poly}). As shown in Fig.~\ref{fig:diff_zone}, the values of $n_{\rm cr}^*$ remain sufficiently close to $n_{\rm cr}$ for hard spectra (illustrated for $\alpha_0=-0.2$), in which case the above factor is of the order of unity, and therefore $S\approx$~const is expected from Eq.~(\ref{eq:flux_conservation}) -- which is indeed seen in the top left panel of Fig.~\ref{fig:output}. On the other hand, for soft spectra ($\alpha_0=2$ in Fig.~\ref{fig:diff_zone}) we have $n_{\rm cr}^*\ll n_{\rm cr}$ due to increasing effect of losses, and hence Eq.~(\ref{eq:flux_conservation}) predicts significant variation of the flux -- which is evident in the top right panel of Fig.~\ref{fig:output}.

\begin{figure}[H]
    \centering
    \includegraphics[width=0.9\linewidth]{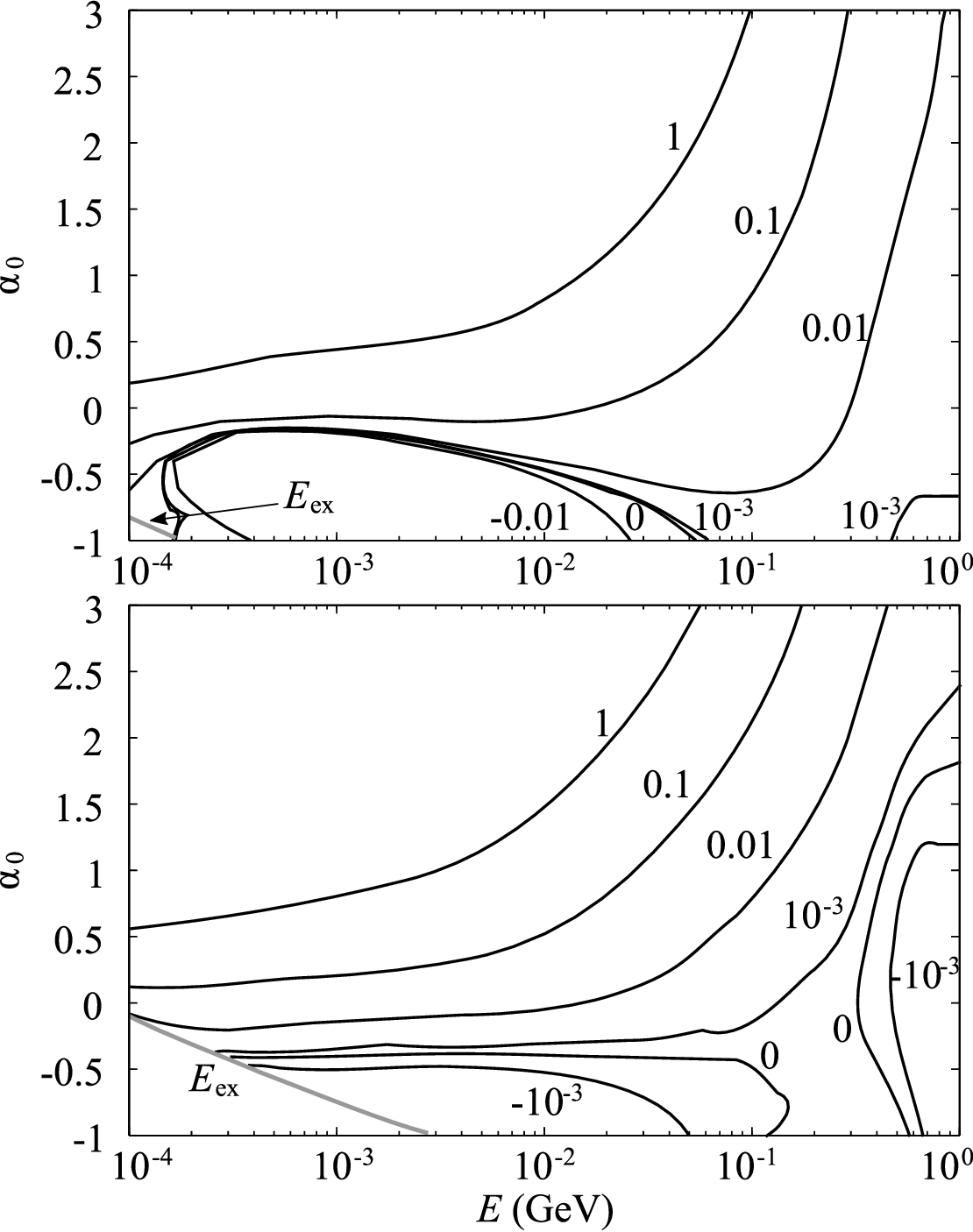}
    \caption{Contour plot showing the relative variation of proton flux across the diffusion zone, $(S_1 - S_2)/S_2$, depending on the kinetic energy $E$ and the spectral index $\alpha_0$ of the power-law distribution (\ref{spectrum_PL}). The top and bottom panels illustrate the effect of  magnitude of CR distribution (see text). The thick gray line indicates the excitation threshold $E_{\rm ex}$, Eq.~(\ref{p_ex}), below which waves cannot be excited.
    }
    \label{fig:delta_S}
\end{figure}

Figure~\ref{fig:delta_S} shows contour lines of the relative flux variation across the diffusion zone, derived from the exact solution for a pure power-law spectrum of interstellar protons, Eq.~(\ref{spectrum_PL}). The results are plotted in the plane of the spectral index and the proton energy. The top panel presents the results for $C_{\rm PL}=C(2m_pc^2/E_{\rm tr})^{2.7-\alpha_0/2}$, where the power-law spectrum matches Eq.~(\ref{spectrum_L}) at lower energies (in particular, coincides with model $\mathscr{L}$ for $\alpha_0=-0.2$), the bottom panel shows the case where $C_{\rm PL}$ is reduced by a factor of 10. We see that the flux is approximately conserved for a fairly broad range of spectral indices around $\alpha_0=-0.3$, corresponding to relatively hard proton spectra (such as model $\mathscr{L}$). This behavior can be qualitatively understood as an approximate balance between CR attenuation operating at a given energy, and replenishment due to losses at higher energies. For very hard spectra, this may even lead to $S_2>S_1$ within a certain range of energies.

\section{Role of adiabatic losses}
\label{app2}

In order to take into account adiabatic losses, we need to add them to regular losses in Eq.~(\ref{eq:main_propagation23}), 
\begin{equation}
    \dot{p}\rightarrow\dot{p}+\frac{p}{3}\frac{dv_{\rm A}}{dz} \,.
\end{equation}
Adiabatic losses become dominant if $\dot{p} \lesssim \frac16v_{\rm A}p(dn/dz)/n$, or
\begin{equation}
    \frac{\Lambda}{\lambda}\lesssim\frac{1+\beta}9 \,.
\label{limit2}
\end{equation}
This condition is stronger than that of Eq.~(\ref{eq:rel_limit}), i.e., adiabatic losses may only dominate in the regime where the effect of regular losses on the diffusion zone is already negligible.

Similar to Appendix~\ref{app1}, now we integrate Eq.~(\ref{eq:main_propagation23}) over $z$ between $z_1$ and $z_2$ neglecting regular losses. Substituting $f \approx (S - S_D)/v_{\rm A}$ into the adiabatic loss term gives 
\begin{equation}
    S_1 - S_2 \approx \frac{1}{6}\int\limits_{n_1}^{n_2}\frac{\partial}{\partial p}\left[p(S - S_D)\right] \frac{dn}{n} =  \frac{\ln (n_2/n_1)}{6} \frac{\partial(pS)}{\partial p} \,.
\end{equation}
As $n_2/n_1 \lesssim 2.5$ and $\partial(pS)/\partial p =\frac34(1-\alpha_0)S$ in the loss-free case [see Eqs.~(22) and (20), respectively, in Ref.~\cite{Chernyshov2024}], we conclude that $|S_1 - S_2| \ll S$ for all reasonable values of the interstellar spectral index, i.e., adiabatic losses have no impact on the flux conservation.

Another effect is associated with the fact that turbulence disappears at the inner boundary, leading to an abrupt drop in the advection velocity from $v_{\rm A2}$ to zero. As a result, adiabatic losses cause a jump in the flux,
\begin{equation}
    S_{2+} - S_{2-} = -\frac{v_{\rm A2}}{3}\frac{\partial (pf_2)}{\partial p} \,,
\end{equation}
so that the flux velocity changes as
\begin{equation}
    u\rightarrow u+\frac{\alpha_2-1}{3}v_{\rm A2} \,.
\end{equation}
Using Eq.~(\ref{eq:u_approximation}) for $u(\mathcal{N})$ we infer that the effect is unimportant if $\mathcal{N}\gtrsim n\lambda~(\propto p^{1+\beta}/\sqrt{n}\,)$. Analysis shows that this condition is always satisfied within the diffusion zones.


\end{appendix}

\bibliographystyle{apsrev4-2}
\bibliography{CR-nonuni}

\begin{thebibliography}{36}%
\makeatletter
\providecommand \@ifxundefined [1]{%
 \@ifx{#1\undefined}
}%
\providecommand \@ifnum [1]{%
 \ifnum #1\expandafter \@firstoftwo
 \else \expandafter \@secondoftwo
 \fi
}%
\providecommand \@ifx [1]{%
 \ifx #1\expandafter \@firstoftwo
 \else \expandafter \@secondoftwo
 \fi
}%
\providecommand \natexlab [1]{#1}%
\providecommand \enquote  [1]{``#1''}%
\providecommand \bibnamefont  [1]{#1}%
\providecommand \bibfnamefont [1]{#1}%
\providecommand \citenamefont [1]{#1}%
\providecommand \href@noop [0]{\@secondoftwo}%
\providecommand \href [0]{\begingroup \@sanitize@url \@href}%
\providecommand \@href[1]{\@@startlink{#1}\@@href}%
\providecommand \@@href[1]{\endgroup#1\@@endlink}%
\providecommand \@sanitize@url [0]{\catcode `\\12\catcode `\$12\catcode
  `\&12\catcode `\#12\catcode `\^12\catcode `\_12\catcode `\%12\relax}%
\providecommand \@@startlink[1]{}%
\providecommand \@@endlink[0]{}%
\providecommand \url  [0]{\begingroup\@sanitize@url \@url }%
\providecommand \@url [1]{\endgroup\@href {#1}{\urlprefix }}%
\providecommand \urlprefix  [0]{URL }%
\providecommand \Eprint [0]{\href }%
\providecommand \doibase [0]{https://doi.org/}%
\providecommand \selectlanguage [0]{\@gobble}%
\providecommand \bibinfo  [0]{\@secondoftwo}%
\providecommand \bibfield  [0]{\@secondoftwo}%
\providecommand \translation [1]{[#1]}%
\providecommand \BibitemOpen [0]{}%
\providecommand \bibitemStop [0]{}%
\providecommand \bibitemNoStop [0]{.\EOS\space}%
\providecommand \EOS [0]{\spacefactor3000\relax}%
\providecommand \BibitemShut  [1]{\csname bibitem#1\endcsname}%
\let\auto@bib@innerbib\@empty
\bibitem [{\citenamefont {{Chernyshov}}\ \emph {et~al.}(2024)\citenamefont
  {{Chernyshov}}, \citenamefont {{Ivlev}},\ and\ \citenamefont
  {{Kiselev}}}]{Chernyshov2024}%
  \BibitemOpen
  \bibfield  {author} {\bibinfo {author} {\bibfnamefont {D.~O.}\ \bibnamefont
  {{Chernyshov}}}, \bibinfo {author} {\bibfnamefont {A.~V.}\ \bibnamefont
  {{Ivlev}}},\ and\ \bibinfo {author} {\bibfnamefont {A.~M.}\ \bibnamefont
  {{Kiselev}}},\ }\href {https://doi.org/10.1103/PhysRevD.110.043012}
  {\bibfield  {journal} {\bibinfo  {journal} {\prd}\ }\textbf {\bibinfo
  {volume} {110}},\ \bibinfo {eid} {043012} (\bibinfo {year}
  {2024})}\BibitemShut {NoStop}%
\bibitem [{\citenamefont {{Skilling}}\ and\ \citenamefont
  {{Strong}}(1976)}]{skill76}%
  \BibitemOpen
  \bibfield  {author} {\bibinfo {author} {\bibfnamefont {J.}~\bibnamefont
  {{Skilling}}}\ and\ \bibinfo {author} {\bibfnamefont {A.~W.}\ \bibnamefont
  {{Strong}}},\ }\href@noop {} {\bibfield  {journal} {\bibinfo  {journal}
  {\aap}\ }\textbf {\bibinfo {volume} {53}},\ \bibinfo {pages} {253} (\bibinfo
  {year} {1976})}\BibitemShut {NoStop}%
\bibitem [{\citenamefont {{Ivlev}}\ \emph {et~al.}(2018)\citenamefont
  {{Ivlev}}, \citenamefont {{Dogiel}}, \citenamefont {{Chernyshov}},
  \citenamefont {{Caselli}}, \citenamefont {{Ko}},\ and\ \citenamefont
  {{Cheng}}}]{ivlev18}%
  \BibitemOpen
  \bibfield  {author} {\bibinfo {author} {\bibfnamefont {A.~V.}\ \bibnamefont
  {{Ivlev}}}, \bibinfo {author} {\bibfnamefont {V.~A.}\ \bibnamefont
  {{Dogiel}}}, \bibinfo {author} {\bibfnamefont {D.~O.}\ \bibnamefont
  {{Chernyshov}}}, \bibinfo {author} {\bibfnamefont {P.}~\bibnamefont
  {{Caselli}}}, \bibinfo {author} {\bibfnamefont {C.~M.}\ \bibnamefont
  {{Ko}}},\ and\ \bibinfo {author} {\bibfnamefont {K.~S.}\ \bibnamefont
  {{Cheng}}},\ }\href {https://doi.org/10.3847/1538-4357/aaadb9} {\bibfield
  {journal} {\bibinfo  {journal} {\apj}\ }\textbf {\bibinfo {volume} {855}},\
  \bibinfo {eid} {23} (\bibinfo {year} {2018})}\BibitemShut {NoStop}%
\bibitem [{\citenamefont {{Dogiel}}\ \emph {et~al.}(2018)\citenamefont
  {{Dogiel}}, \citenamefont {{Chernyshov}}, \citenamefont {{Ivlev}},
  \citenamefont {{Malyshev}}, \citenamefont {{Strong}},\ and\ \citenamefont
  {{Cheng}}}]{dogiel2018}%
  \BibitemOpen
  \bibfield  {author} {\bibinfo {author} {\bibfnamefont {V.~A.}\ \bibnamefont
  {{Dogiel}}}, \bibinfo {author} {\bibfnamefont {D.~O.}\ \bibnamefont
  {{Chernyshov}}}, \bibinfo {author} {\bibfnamefont {A.~V.}\ \bibnamefont
  {{Ivlev}}}, \bibinfo {author} {\bibfnamefont {D.}~\bibnamefont {{Malyshev}}},
  \bibinfo {author} {\bibfnamefont {A.~W.}\ \bibnamefont {{Strong}}},\ and\
  \bibinfo {author} {\bibfnamefont {K.~S.}\ \bibnamefont {{Cheng}}},\ }\href
  {https://doi.org/10.3847/1538-4357/aae827} {\bibfield  {journal} {\bibinfo
  {journal} {\apj}\ }\textbf {\bibinfo {volume} {868}},\ \bibinfo {eid} {114}
  (\bibinfo {year} {2018})}\BibitemShut {NoStop}%
\bibitem [{\citenamefont {{Kulsrud}}\ and\ \citenamefont
  {{Pearce}}(1969)}]{kuls69}%
  \BibitemOpen
  \bibfield  {author} {\bibinfo {author} {\bibfnamefont {R.}~\bibnamefont
  {{Kulsrud}}}\ and\ \bibinfo {author} {\bibfnamefont {W.~P.}\ \bibnamefont
  {{Pearce}}},\ }\href {https://doi.org/10.1086/149981} {\bibfield  {journal}
  {\bibinfo  {journal} {\apj}\ }\textbf {\bibinfo {volume} {156}},\ \bibinfo
  {pages} {445} (\bibinfo {year} {1969})}\BibitemShut {NoStop}%
\bibitem [{\citenamefont {{Yang}}\ \emph {et~al.}(2023)\citenamefont {{Yang}},
  \citenamefont {{Li}}, \citenamefont {{Wilhelmi}}, \citenamefont {{Cui}},
  \citenamefont {{Liu}},\ and\ \citenamefont {{Aharonian}}}]{yang2023}%
  \BibitemOpen
  \bibfield  {author} {\bibinfo {author} {\bibfnamefont {R.-z.}\ \bibnamefont
  {{Yang}}}, \bibinfo {author} {\bibfnamefont {G.-X.}\ \bibnamefont {{Li}}},
  \bibinfo {author} {\bibfnamefont {E.~d.~O.}\ \bibnamefont {{Wilhelmi}}},
  \bibinfo {author} {\bibfnamefont {Y.-D.}\ \bibnamefont {{Cui}}}, \bibinfo
  {author} {\bibfnamefont {B.}~\bibnamefont {{Liu}}},\ and\ \bibinfo {author}
  {\bibfnamefont {F.}~\bibnamefont {{Aharonian}}},\ }\href
  {https://doi.org/10.1038/s41550-022-01868-9} {\bibfield  {journal} {\bibinfo
  {journal} {Nature Astronomy}\ }\textbf {\bibinfo {volume} {7}},\ \bibinfo
  {pages} {351} (\bibinfo {year} {2023})}\BibitemShut {NoStop}%
\bibitem [{\citenamefont {{Padovani}}\ \emph {et~al.}(2020)\citenamefont
  {{Padovani}}, \citenamefont {{Ivlev}}, \citenamefont {{Galli}}, \citenamefont
  {{Offner}}, \citenamefont {{Indriolo}}, \citenamefont {{Rodgers-Lee}},
  \citenamefont {{Marcowith}}, \citenamefont {{Girichidis}}, \citenamefont
  {{Bykov}},\ and\ \citenamefont {{Kruijssen}}}]{Padovani2020}%
  \BibitemOpen
  \bibfield  {author} {\bibinfo {author} {\bibfnamefont {M.}~\bibnamefont
  {{Padovani}}}, \bibinfo {author} {\bibfnamefont {A.~V.}\ \bibnamefont
  {{Ivlev}}}, \bibinfo {author} {\bibfnamefont {D.}~\bibnamefont {{Galli}}},
  \bibinfo {author} {\bibfnamefont {S.~S.~R.}\ \bibnamefont {{Offner}}},
  \bibinfo {author} {\bibfnamefont {N.}~\bibnamefont {{Indriolo}}}, \bibinfo
  {author} {\bibfnamefont {D.}~\bibnamefont {{Rodgers-Lee}}}, \bibinfo {author}
  {\bibfnamefont {A.}~\bibnamefont {{Marcowith}}}, \bibinfo {author}
  {\bibfnamefont {P.}~\bibnamefont {{Girichidis}}}, \bibinfo {author}
  {\bibfnamefont {A.~M.}\ \bibnamefont {{Bykov}}},\ and\ \bibinfo {author}
  {\bibfnamefont {J.~M.~D.}\ \bibnamefont {{Kruijssen}}},\ }\href
  {https://doi.org/10.1007/s11214-020-00654-1} {\bibfield  {journal} {\bibinfo
  {journal} {\ssr}\ }\textbf {\bibinfo {volume} {216}},\ \bibinfo {eid} {29}
  (\bibinfo {year} {2020})}\BibitemShut {NoStop}%
\bibitem [{\citenamefont {{Gabici}}(2022)}]{Gabici2022}%
  \BibitemOpen
  \bibfield  {author} {\bibinfo {author} {\bibfnamefont {S.}~\bibnamefont
  {{Gabici}}},\ }\href {https://doi.org/10.1007/s00159-022-00141-2} {\bibfield
  {journal} {\bibinfo  {journal} {\aapr}\ }\textbf {\bibinfo {volume} {30}},\
  \bibinfo {eid} {4} (\bibinfo {year} {2022})}\BibitemShut {NoStop}%
\bibitem [{\citenamefont {{Ramaty}}\ \emph {et~al.}(1979)\citenamefont
  {{Ramaty}}, \citenamefont {{Kozlovsky}},\ and\ \citenamefont
  {{Lingenfelter}}}]{Ramaty79}%
  \BibitemOpen
  \bibfield  {author} {\bibinfo {author} {\bibfnamefont {R.}~\bibnamefont
  {{Ramaty}}}, \bibinfo {author} {\bibfnamefont {B.}~\bibnamefont
  {{Kozlovsky}}},\ and\ \bibinfo {author} {\bibfnamefont {R.~E.}\ \bibnamefont
  {{Lingenfelter}}},\ }\href {https://doi.org/10.1086/190596} {\bibfield
  {journal} {\bibinfo  {journal} {\apjs}\ }\textbf {\bibinfo {volume} {40}},\
  \bibinfo {pages} {487} (\bibinfo {year} {1979})}\BibitemShut {NoStop}%
\bibitem [{\citenamefont {{Tatischeff}}(2003)}]{Tatischeff03}%
  \BibitemOpen
  \bibfield  {author} {\bibinfo {author} {\bibfnamefont {V.}~\bibnamefont
  {{Tatischeff}}},\ }in\ \href {https://doi.org/10.1051/eas:2003038} {\emph
  {\bibinfo {booktitle} {EAS Publications Series}}},\ \bibinfo {series} {EAS
  Publications Series}, Vol.~\bibinfo {volume} {7},\ \bibinfo {editor} {edited
  by\ \bibinfo {editor} {\bibfnamefont {C.}~\bibnamefont {{Motch}}}\ and\
  \bibinfo {editor} {\bibfnamefont {J.-M.}\ \bibnamefont {{Hameury}}}}\
  (\bibinfo {year} {2003})\ p.~\bibinfo {pages} {79}\BibitemShut {NoStop}%
\bibitem [{\citenamefont {{Benhabiles-Mezhoud}}\ \emph
  {et~al.}(2013)\citenamefont {{Benhabiles-Mezhoud}}, \citenamefont {{Kiener}},
  \citenamefont {{Tatischeff}},\ and\ \citenamefont
  {{Strong}}}]{Benhabiles2013}%
  \BibitemOpen
  \bibfield  {author} {\bibinfo {author} {\bibfnamefont {H.}~\bibnamefont
  {{Benhabiles-Mezhoud}}}, \bibinfo {author} {\bibfnamefont {J.}~\bibnamefont
  {{Kiener}}}, \bibinfo {author} {\bibfnamefont {V.}~\bibnamefont
  {{Tatischeff}}},\ and\ \bibinfo {author} {\bibfnamefont {A.~W.}\ \bibnamefont
  {{Strong}}},\ }\href {https://doi.org/10.1088/0004-637X/766/2/139} {\bibfield
   {journal} {\bibinfo  {journal} {\apj}\ }\textbf {\bibinfo {volume} {766}},\
  \bibinfo {eid} {139} (\bibinfo {year} {2013})}\BibitemShut {NoStop}%
\bibitem [{\citenamefont {{Phan}}\ \emph {et~al.}(2020)\citenamefont {{Phan}},
  \citenamefont {{Gabici}}, \citenamefont {{Morlino}}, \citenamefont
  {{Terrier}}, \citenamefont {{Vink}}, \citenamefont {{Krause}},\ and\
  \citenamefont {{Menu}}}]{Phan2020}%
  \BibitemOpen
  \bibfield  {author} {\bibinfo {author} {\bibfnamefont {V.~H.~M.}\
  \bibnamefont {{Phan}}}, \bibinfo {author} {\bibfnamefont {S.}~\bibnamefont
  {{Gabici}}}, \bibinfo {author} {\bibfnamefont {G.}~\bibnamefont {{Morlino}}},
  \bibinfo {author} {\bibfnamefont {R.}~\bibnamefont {{Terrier}}}, \bibinfo
  {author} {\bibfnamefont {J.}~\bibnamefont {{Vink}}}, \bibinfo {author}
  {\bibfnamefont {J.}~\bibnamefont {{Krause}}},\ and\ \bibinfo {author}
  {\bibfnamefont {M.}~\bibnamefont {{Menu}}},\ }\href
  {https://doi.org/10.1051/0004-6361/201936927} {\bibfield  {journal} {\bibinfo
   {journal} {\aap}\ }\textbf {\bibinfo {volume} {635}},\ \bibinfo {eid} {A40}
  (\bibinfo {year} {2020})}\BibitemShut {NoStop}%
\bibitem [{\citenamefont {{Fujita}}\ \emph {et~al.}(2021)\citenamefont
  {{Fujita}}, \citenamefont {{Nobukawa}},\ and\ \citenamefont
  {{Sano}}}]{Fujita2021}%
  \BibitemOpen
  \bibfield  {author} {\bibinfo {author} {\bibfnamefont {Y.}~\bibnamefont
  {{Fujita}}}, \bibinfo {author} {\bibfnamefont {K.~K.}\ \bibnamefont
  {{Nobukawa}}},\ and\ \bibinfo {author} {\bibfnamefont {H.}~\bibnamefont
  {{Sano}}},\ }\href {https://doi.org/10.3847/1538-4357/abce62} {\bibfield
  {journal} {\bibinfo  {journal} {\apj}\ }\textbf {\bibinfo {volume} {908}},\
  \bibinfo {eid} {136} (\bibinfo {year} {2021})}\BibitemShut {NoStop}%
\bibitem [{\citenamefont {{Liu}}\ \emph {et~al.}(2021)\citenamefont {{Liu}},
  \citenamefont {{Yang}},\ and\ \citenamefont {{Aharonian}}}]{Liu2021}%
  \BibitemOpen
  \bibfield  {author} {\bibinfo {author} {\bibfnamefont {B.}~\bibnamefont
  {{Liu}}}, \bibinfo {author} {\bibfnamefont {R.-z.}\ \bibnamefont {{Yang}}},\
  and\ \bibinfo {author} {\bibfnamefont {F.}~\bibnamefont {{Aharonian}}},\
  }\href {https://doi.org/10.1051/0004-6361/202039977} {\bibfield  {journal}
  {\bibinfo  {journal} {\aap}\ }\textbf {\bibinfo {volume} {646}},\ \bibinfo
  {eid} {A149} (\bibinfo {year} {2021})}\BibitemShut {NoStop}%
\bibitem [{\citenamefont {{Shi}}\ \emph {et~al.}(2024)\citenamefont {{Shi}},
  \citenamefont {{Liu}},\ and\ \citenamefont {{Yang}}}]{shi2024}%
  \BibitemOpen
  \bibfield  {author} {\bibinfo {author} {\bibfnamefont {Z.}~\bibnamefont
  {{Shi}}}, \bibinfo {author} {\bibfnamefont {B.}~\bibnamefont {{Liu}}},\ and\
  \bibinfo {author} {\bibfnamefont {R.-z.}\ \bibnamefont {{Yang}}},\ }\href
  {https://doi.org/10.1051/0004-6361/202451973} {\bibfield  {journal} {\bibinfo
   {journal} {\aap}\ }\textbf {\bibinfo {volume} {692}},\ \bibinfo {eid} {A128}
  (\bibinfo {year} {2024})}\BibitemShut {NoStop}%
\bibitem [{\citenamefont {{Padovani}}\ \emph {et~al.}(2018)\citenamefont
  {{Padovani}}, \citenamefont {{Ivlev}}, \citenamefont {{Galli}},\ and\
  \citenamefont {{Caselli}}}]{Padovani2018}%
  \BibitemOpen
  \bibfield  {author} {\bibinfo {author} {\bibfnamefont {M.}~\bibnamefont
  {{Padovani}}}, \bibinfo {author} {\bibfnamefont {A.~V.}\ \bibnamefont
  {{Ivlev}}}, \bibinfo {author} {\bibfnamefont {D.}~\bibnamefont {{Galli}}},\
  and\ \bibinfo {author} {\bibfnamefont {P.}~\bibnamefont {{Caselli}}},\ }\href
  {https://doi.org/10.1051/0004-6361/201732202} {\bibfield  {journal} {\bibinfo
   {journal} {\aap}\ }\textbf {\bibinfo {volume} {614}},\ \bibinfo {eid} {A111}
  (\bibinfo {year} {2018})}\BibitemShut {NoStop}%
\bibitem [{\citenamefont {{Berezinskii}}\ \emph {et~al.}(1990)\citenamefont
  {{Berezinskii}}, \citenamefont {{Bulanov}}, \citenamefont {{Dogiel}},
  \citenamefont {{Ginzburg}},\ and\ \citenamefont
  {{Ptuskin}}}]{BerezinskiiBook1990}%
  \BibitemOpen
  \bibfield  {author} {\bibinfo {author} {\bibfnamefont {V.~S.}\ \bibnamefont
  {{Berezinskii}}}, \bibinfo {author} {\bibfnamefont {S.~V.}\ \bibnamefont
  {{Bulanov}}}, \bibinfo {author} {\bibfnamefont {V.~A.}\ \bibnamefont
  {{Dogiel}}}, \bibinfo {author} {\bibfnamefont {V.~L.}\ \bibnamefont
  {{Ginzburg}}},\ and\ \bibinfo {author} {\bibfnamefont {V.~S.}\ \bibnamefont
  {{Ptuskin}}},\ }\href@noop {} {\emph {\bibinfo {title} {{Astrophysics of
  cosmic rays}}}}\ (\bibinfo  {publisher} {Amsterdam: North-Holland},\ \bibinfo
  {year} {1990})\BibitemShut {NoStop}%
\bibitem [{\citenamefont {{Skilling}}(1975)}]{Skilling1975}%
  \BibitemOpen
  \bibfield  {author} {\bibinfo {author} {\bibfnamefont {J.}~\bibnamefont
  {{Skilling}}},\ }\href {https://doi.org/10.1093/mnras/173.2.255} {\bibfield
  {journal} {\bibinfo  {journal} {\mnras}\ }\textbf {\bibinfo {volume} {173}},\
  \bibinfo {pages} {255} (\bibinfo {year} {1975})}\BibitemShut {NoStop}%
\bibitem [{\citenamefont {{Padovani}}\ \emph {et~al.}(2009)\citenamefont
  {{Padovani}}, \citenamefont {{Galli}},\ and\ \citenamefont
  {{Glassgold}}}]{Padovani2009}%
  \BibitemOpen
  \bibfield  {author} {\bibinfo {author} {\bibfnamefont {M.}~\bibnamefont
  {{Padovani}}}, \bibinfo {author} {\bibfnamefont {D.}~\bibnamefont
  {{Galli}}},\ and\ \bibinfo {author} {\bibfnamefont {A.~E.}\ \bibnamefont
  {{Glassgold}}},\ }\href {https://doi.org/10.1051/0004-6361/200911794}
  {\bibfield  {journal} {\bibinfo  {journal} {\aap}\ }\textbf {\bibinfo
  {volume} {501}},\ \bibinfo {pages} {619} (\bibinfo {year}
  {2009})}\BibitemShut {NoStop}%
\bibitem [{\citenamefont {{Silsbee}}\ and\ \citenamefont
  {{Ivlev}}(2019)}]{Silsbee2019}%
  \BibitemOpen
  \bibfield  {author} {\bibinfo {author} {\bibfnamefont {K.}~\bibnamefont
  {{Silsbee}}}\ and\ \bibinfo {author} {\bibfnamefont {A.~V.}\ \bibnamefont
  {{Ivlev}}},\ }\href {https://doi.org/10.3847/1538-4357/ab22b4} {\bibfield
  {journal} {\bibinfo  {journal} {\apj}\ }\textbf {\bibinfo {volume} {879}},\
  \bibinfo {eid} {14} (\bibinfo {year} {2019})}\BibitemShut {NoStop}%
\bibitem [{\citenamefont {{Crutcher}}(2012)}]{Crutcher2012}%
  \BibitemOpen
  \bibfield  {author} {\bibinfo {author} {\bibfnamefont {R.~M.}\ \bibnamefont
  {{Crutcher}}},\ }\href {https://doi.org/10.1146/annurev-astro-081811-125514}
  {\bibfield  {journal} {\bibinfo  {journal} {\araa}\ }\textbf {\bibinfo
  {volume} {50}},\ \bibinfo {pages} {29} (\bibinfo {year} {2012})}\BibitemShut
  {NoStop}%
\bibitem [{\citenamefont {{Edenhofer}}\ \emph {et~al.}(2024)\citenamefont
  {{Edenhofer}}, \citenamefont {{Zucker}}, \citenamefont {{Frank}},
  \citenamefont {{Saydjari}}, \citenamefont {{Speagle}}, \citenamefont
  {{Finkbeiner}},\ and\ \citenamefont {{En{\ss}lin}}}]{Edenhofer2024}%
  \BibitemOpen
  \bibfield  {author} {\bibinfo {author} {\bibfnamefont {G.}~\bibnamefont
  {{Edenhofer}}}, \bibinfo {author} {\bibfnamefont {C.}~\bibnamefont
  {{Zucker}}}, \bibinfo {author} {\bibfnamefont {P.}~\bibnamefont {{Frank}}},
  \bibinfo {author} {\bibfnamefont {A.~K.}\ \bibnamefont {{Saydjari}}},
  \bibinfo {author} {\bibfnamefont {J.~S.}\ \bibnamefont {{Speagle}}}, \bibinfo
  {author} {\bibfnamefont {D.}~\bibnamefont {{Finkbeiner}}},\ and\ \bibinfo
  {author} {\bibfnamefont {T.~A.}\ \bibnamefont {{En{\ss}lin}}},\ }\href
  {https://doi.org/10.1051/0004-6361/202347628} {\bibfield  {journal} {\bibinfo
   {journal} {\aap}\ }\textbf {\bibinfo {volume} {685}},\ \bibinfo {eid} {A82}
  (\bibinfo {year} {2024})}\BibitemShut {NoStop}%
\bibitem [{\citenamefont {{Obolentseva}}\ \emph {et~al.}(2024)\citenamefont
  {{Obolentseva}}, \citenamefont {{Ivlev}}, \citenamefont {{Silsbee}},
  \citenamefont {{Neufeld}}, \citenamefont {{Caselli}}, \citenamefont
  {{Edenhofer}}, \citenamefont {{Indriolo}}, \citenamefont {{Bisbas}},\ and\
  \citenamefont {{Lomeli}}}]{Obolentseva2024}%
  \BibitemOpen
  \bibfield  {author} {\bibinfo {author} {\bibfnamefont {M.}~\bibnamefont
  {{Obolentseva}}}, \bibinfo {author} {\bibfnamefont {A.~V.}\ \bibnamefont
  {{Ivlev}}}, \bibinfo {author} {\bibfnamefont {K.}~\bibnamefont {{Silsbee}}},
  \bibinfo {author} {\bibfnamefont {D.~A.}\ \bibnamefont {{Neufeld}}}, \bibinfo
  {author} {\bibfnamefont {P.}~\bibnamefont {{Caselli}}}, \bibinfo {author}
  {\bibfnamefont {G.}~\bibnamefont {{Edenhofer}}}, \bibinfo {author}
  {\bibfnamefont {N.}~\bibnamefont {{Indriolo}}}, \bibinfo {author}
  {\bibfnamefont {T.~G.}\ \bibnamefont {{Bisbas}}},\ and\ \bibinfo {author}
  {\bibfnamefont {D.}~\bibnamefont {{Lomeli}}},\ }\href
  {https://doi.org/10.3847/1538-4357/ad71ce} {\bibfield  {journal} {\bibinfo
  {journal} {\apj}\ }\textbf {\bibinfo {volume} {973}},\ \bibinfo {eid} {142}
  (\bibinfo {year} {2024})}\BibitemShut {NoStop}%
\bibitem [{\citenamefont {{Aguilar}}\ \emph {et~al.}(2021)\citenamefont
  {{Aguilar}}, \citenamefont {{Ali Cavasonza}}, \citenamefont {{Ambrosi}},
  \citenamefont {{Arruda}}, \citenamefont {{Attig}}, \citenamefont {{Barao}},
  \citenamefont {{Barrin}}, \citenamefont {{Bartoloni}}, \citenamefont
  {{Ba{\c{s}}e{\u{g}}mez-du Pree}}, \citenamefont {{Bates}},\ and\
  \citenamefont {et~al.}}]{Aguilar2021}%
  \BibitemOpen
  \bibfield  {author} {\bibinfo {author} {\bibfnamefont {M.}~\bibnamefont
  {{Aguilar}}}, \bibinfo {author} {\bibfnamefont {L.}~\bibnamefont {{Ali
  Cavasonza}}}, \bibinfo {author} {\bibfnamefont {G.}~\bibnamefont
  {{Ambrosi}}}, \bibinfo {author} {\bibfnamefont {L.}~\bibnamefont {{Arruda}}},
  \bibinfo {author} {\bibfnamefont {N.}~\bibnamefont {{Attig}}}, \bibinfo
  {author} {\bibfnamefont {F.}~\bibnamefont {{Barao}}}, \bibinfo {author}
  {\bibfnamefont {L.}~\bibnamefont {{Barrin}}}, \bibinfo {author}
  {\bibfnamefont {A.}~\bibnamefont {{Bartoloni}}}, \bibinfo {author}
  {\bibfnamefont {S.}~\bibnamefont {{Ba{\c{s}}e{\u{g}}mez-du Pree}}}, \bibinfo
  {author} {\bibfnamefont {J.}~\bibnamefont {{Bates}}},\ and\ \bibinfo {author}
  {\bibnamefont {et~al.}},\ }\href
  {https://doi.org/10.1016/j.physrep.2020.09.003} {\bibfield  {journal}
  {\bibinfo  {journal} {\physrep}\ }\textbf {\bibinfo {volume} {894}},\
  \bibinfo {pages} {1} (\bibinfo {year} {2021})}\BibitemShut {NoStop}%
\bibitem [{\citenamefont {{Cummings}}\ \emph {et~al.}(2016)\citenamefont
  {{Cummings}}, \citenamefont {{Stone}}, \citenamefont {{Heikkila}},
  \citenamefont {{Lal}}, \citenamefont {{Webber}}, \citenamefont
  {{J{\'o}hannesson}}, \citenamefont {{Moskalenko}}, \citenamefont
  {{Orlando}},\ and\ \citenamefont {{Porter}}}]{Cummings2016}%
  \BibitemOpen
  \bibfield  {author} {\bibinfo {author} {\bibfnamefont {A.~C.}\ \bibnamefont
  {{Cummings}}}, \bibinfo {author} {\bibfnamefont {E.~C.}\ \bibnamefont
  {{Stone}}}, \bibinfo {author} {\bibfnamefont {B.~C.}\ \bibnamefont
  {{Heikkila}}}, \bibinfo {author} {\bibfnamefont {N.}~\bibnamefont {{Lal}}},
  \bibinfo {author} {\bibfnamefont {W.~R.}\ \bibnamefont {{Webber}}}, \bibinfo
  {author} {\bibfnamefont {G.}~\bibnamefont {{J{\'o}hannesson}}}, \bibinfo
  {author} {\bibfnamefont {I.~V.}\ \bibnamefont {{Moskalenko}}}, \bibinfo
  {author} {\bibfnamefont {E.}~\bibnamefont {{Orlando}}},\ and\ \bibinfo
  {author} {\bibfnamefont {T.~A.}\ \bibnamefont {{Porter}}},\ }\href
  {https://doi.org/10.3847/0004-637X/831/1/18} {\bibfield  {journal} {\bibinfo
  {journal} {\apj}\ }\textbf {\bibinfo {volume} {831}},\ \bibinfo {eid} {18}
  (\bibinfo {year} {2016})}\BibitemShut {NoStop}%
\bibitem [{\citenamefont {{Stone}}\ \emph {et~al.}(2019)\citenamefont
  {{Stone}}, \citenamefont {{Cummings}}, \citenamefont {{Heikkila}},\ and\
  \citenamefont {{Lal}}}]{Stone2019}%
  \BibitemOpen
  \bibfield  {author} {\bibinfo {author} {\bibfnamefont {E.~C.}\ \bibnamefont
  {{Stone}}}, \bibinfo {author} {\bibfnamefont {A.~C.}\ \bibnamefont
  {{Cummings}}}, \bibinfo {author} {\bibfnamefont {B.~C.}\ \bibnamefont
  {{Heikkila}}},\ and\ \bibinfo {author} {\bibfnamefont {N.}~\bibnamefont
  {{Lal}}},\ }\href {https://doi.org/10.1038/s41550-019-0928-3} {\bibfield
  {journal} {\bibinfo  {journal} {Nature Astronomy}\ }\textbf {\bibinfo
  {volume} {3}},\ \bibinfo {pages} {1013} (\bibinfo {year} {2019})}\BibitemShut
  {NoStop}%
\bibitem [{\citenamefont {{Ginzburg}}(1979)}]{GinzburgBook1979}%
  \BibitemOpen
  \bibfield  {author} {\bibinfo {author} {\bibfnamefont {V.~L.}\ \bibnamefont
  {{Ginzburg}}},\ }\href@noop {} {\emph {\bibinfo {title} {Theoretical physics
  and astrophysics}}}\ (\bibinfo  {publisher} {Oxford: Pergamon},\ \bibinfo
  {year} {1979})\BibitemShut {NoStop}%
\bibitem [{\citenamefont {{Ginzburg}}(1970)}]{GinzburgBook1970}%
  \BibitemOpen
  \bibfield  {author} {\bibinfo {author} {\bibfnamefont {V.~L.}\ \bibnamefont
  {{Ginzburg}}},\ }\href@noop {} {\emph {\bibinfo {title} {{The propagation of
  electromagnetic waves in plasmas}}}}\ (\bibinfo  {publisher} {Oxford:
  Pergamon},\ \bibinfo {year} {1970})\BibitemShut {NoStop}%
\bibitem [{\citenamefont {{Kulsrud}}(2005)}]{KulsrudBook2005}%
  \BibitemOpen
  \bibfield  {author} {\bibinfo {author} {\bibfnamefont {R.~M.}\ \bibnamefont
  {{Kulsrud}}},\ }\href@noop {} {\emph {\bibinfo {title} {{Plasma Physics for
  Astrophysics}}}}\ (\bibinfo  {publisher} {Princeton, NJ: Princeton Univ.
  Press},\ \bibinfo {year} {2005})\BibitemShut {NoStop}%
\bibitem [{\citenamefont {{Draine}}(2011)}]{DraineBook2011}%
  \BibitemOpen
  \bibfield  {author} {\bibinfo {author} {\bibfnamefont {B.~T.}\ \bibnamefont
  {{Draine}}},\ }\href@noop {} {\emph {\bibinfo {title} {{Physics of the
  Interstellar and Intergalactic Medium}}}}\ (\bibinfo  {publisher} {Princeton,
  NJ: Princeton Univ. Press},\ \bibinfo {year} {2011})\BibitemShut {NoStop}%
\bibitem [{\citenamefont {{Becker}}\ \emph {et~al.}(2011)\citenamefont
  {{Becker}}, \citenamefont {{Black}}, \citenamefont {{Safarzadeh}},\ and\
  \citenamefont {{Schuppan}}}]{Becker2011}%
  \BibitemOpen
  \bibfield  {author} {\bibinfo {author} {\bibfnamefont {J.~K.}\ \bibnamefont
  {{Becker}}}, \bibinfo {author} {\bibfnamefont {J.~H.}\ \bibnamefont
  {{Black}}}, \bibinfo {author} {\bibfnamefont {M.}~\bibnamefont
  {{Safarzadeh}}},\ and\ \bibinfo {author} {\bibfnamefont {F.}~\bibnamefont
  {{Schuppan}}},\ }\href {https://doi.org/10.1088/2041-8205/739/2/L43}
  {\bibfield  {journal} {\bibinfo  {journal} {\apjl}\ }\textbf {\bibinfo
  {volume} {739}},\ \bibinfo {eid} {L43} (\bibinfo {year} {2011})}\BibitemShut
  {NoStop}%
\bibitem [{\citenamefont {{Vaupr{\'e}}}\ \emph {et~al.}(2014)\citenamefont
  {{Vaupr{\'e}}}, \citenamefont {{Hily-Blant}}, \citenamefont {{Ceccarelli}},
  \citenamefont {{Dubus}}, \citenamefont {{Gabici}},\ and\ \citenamefont
  {{Montmerle}}}]{Vaupre2014}%
  \BibitemOpen
  \bibfield  {author} {\bibinfo {author} {\bibfnamefont {S.}~\bibnamefont
  {{Vaupr{\'e}}}}, \bibinfo {author} {\bibfnamefont {P.}~\bibnamefont
  {{Hily-Blant}}}, \bibinfo {author} {\bibfnamefont {C.}~\bibnamefont
  {{Ceccarelli}}}, \bibinfo {author} {\bibfnamefont {G.}~\bibnamefont
  {{Dubus}}}, \bibinfo {author} {\bibfnamefont {S.}~\bibnamefont {{Gabici}}},\
  and\ \bibinfo {author} {\bibfnamefont {T.}~\bibnamefont {{Montmerle}}},\
  }\href {https://doi.org/10.1051/0004-6361/201424036} {\bibfield  {journal}
  {\bibinfo  {journal} {\aap}\ }\textbf {\bibinfo {volume} {568}},\ \bibinfo
  {eid} {A50} (\bibinfo {year} {2014})}\BibitemShut {NoStop}%
\bibitem [{\citenamefont {{Owen}}\ \emph {et~al.}(2021)\citenamefont {{Owen}},
  \citenamefont {{On}}, \citenamefont {{Lai}},\ and\ \citenamefont
  {{Wu}}}]{Owen2021}%
  \BibitemOpen
  \bibfield  {author} {\bibinfo {author} {\bibfnamefont {E.~R.}\ \bibnamefont
  {{Owen}}}, \bibinfo {author} {\bibfnamefont {A.~Y.~L.}\ \bibnamefont {{On}}},
  \bibinfo {author} {\bibfnamefont {S.-P.}\ \bibnamefont {{Lai}}},\ and\
  \bibinfo {author} {\bibfnamefont {K.}~\bibnamefont {{Wu}}},\ }\href
  {https://doi.org/10.3847/1538-4357/abee1a} {\bibfield  {journal} {\bibinfo
  {journal} {\apj}\ }\textbf {\bibinfo {volume} {913}},\ \bibinfo {eid} {52}
  (\bibinfo {year} {2021})}\BibitemShut {NoStop}%
\bibitem [{\citenamefont {{Dogiel}}\ \emph {et~al.}(2015)\citenamefont
  {{Dogiel}}, \citenamefont {{Chernyshov}}, \citenamefont {{Kiselev}},
  \citenamefont {{Nobukawa}}, \citenamefont {{Cheng}}, \citenamefont {{Hui}},
  \citenamefont {{Ko}}, \citenamefont {{Nobukawa}},\ and\ \citenamefont
  {{Tsuru}}}]{dogiel2015}%
  \BibitemOpen
  \bibfield  {author} {\bibinfo {author} {\bibfnamefont {V.~A.}\ \bibnamefont
  {{Dogiel}}}, \bibinfo {author} {\bibfnamefont {D.~O.}\ \bibnamefont
  {{Chernyshov}}}, \bibinfo {author} {\bibfnamefont {A.~M.}\ \bibnamefont
  {{Kiselev}}}, \bibinfo {author} {\bibfnamefont {M.}~\bibnamefont
  {{Nobukawa}}}, \bibinfo {author} {\bibfnamefont {K.~S.}\ \bibnamefont
  {{Cheng}}}, \bibinfo {author} {\bibfnamefont {C.~Y.}\ \bibnamefont {{Hui}}},
  \bibinfo {author} {\bibfnamefont {C.~M.}\ \bibnamefont {{Ko}}}, \bibinfo
  {author} {\bibfnamefont {K.~K.}\ \bibnamefont {{Nobukawa}}},\ and\ \bibinfo
  {author} {\bibfnamefont {T.~G.}\ \bibnamefont {{Tsuru}}},\ }\href
  {https://doi.org/10.1088/0004-637X/809/1/48} {\bibfield  {journal} {\bibinfo
  {journal} {\apj}\ }\textbf {\bibinfo {volume} {809}},\ \bibinfo {eid} {48}
  (\bibinfo {year} {2015})}\BibitemShut {NoStop}%
\bibitem [{\citenamefont {{Ravikularaman}}\ \emph {et~al.}(2025)\citenamefont
  {{Ravikularaman}}, \citenamefont {{Recchia}}, \citenamefont {{Phan}},\ and\
  \citenamefont {{Gabici}}}]{Ravikularaman2025}%
  \BibitemOpen
  \bibfield  {author} {\bibinfo {author} {\bibfnamefont {S.}~\bibnamefont
  {{Ravikularaman}}}, \bibinfo {author} {\bibfnamefont {S.}~\bibnamefont
  {{Recchia}}}, \bibinfo {author} {\bibfnamefont {V.~H.~M.}\ \bibnamefont
  {{Phan}}},\ and\ \bibinfo {author} {\bibfnamefont {S.}~\bibnamefont
  {{Gabici}}},\ }\href {https://doi.org/10.1051/0004-6361/202451155} {\bibfield
   {journal} {\bibinfo  {journal} {\aap}\ }\textbf {\bibinfo {volume} {694}},\
  \bibinfo {eid} {A114} (\bibinfo {year} {2025})}\BibitemShut {NoStop}%
\bibitem [{\citenamefont {{Indriolo}}\ and\ \citenamefont
  {{McCall}}(2013)}]{Indriolo2013}%
  \BibitemOpen
  \bibfield  {author} {\bibinfo {author} {\bibfnamefont {N.}~\bibnamefont
  {{Indriolo}}}\ and\ \bibinfo {author} {\bibfnamefont {B.~J.}\ \bibnamefont
  {{McCall}}},\ }\href {https://doi.org/10.1039/C3CS60087D} {\bibfield
  {journal} {\bibinfo  {journal} {\csr}\ }\textbf {\bibinfo {volume} {42}},\
  \bibinfo {pages} {7763} (\bibinfo {year} {2013})}\BibitemShut {NoStop}%
\end{thebibliography}%

\end{document}